\documentclass[11pt]{article}
\usepackage{amssymb,amsmath,amsfonts}
\usepackage{graphicx}
\usepackage{graphics}
\usepackage{eepic,epsfig}
\usepackage{verbatim}
\usepackage{color}
\usepackage{hyperref}
1.60cm \makeatletter \@addtoreset{equation}{section}

\makeatother

\begin{document}
\title{Vacuum polarization by charged  matter fields in AdS spacetime in the presence of a cosmic string}
\author{ E. R. Bezerra de Mello\thanks
	{E-mail: emello@fisica.ufpb.br}.\\
\\
\textit{Departamento de F\'{\i}sica, Universidade Federal da Para\'{\i}ba}\\
\textit{58059-900, Caixa Postal 5008, Jo\~{a}o Pessoa, PB, Brazil.}\vspace{%
0.3cm}\\
}
\maketitle
%
\begin{abstract}
In this paper we revisited the analyse the vacuum polarizations effect associated wit charged matter fields in a higher-dimensional anti de-Sitter (AdS) spacetime in the presence of an idealized cosmic string carrying  magnetic flux running along its core. Specifically we will consider, separately, charged bosonic and fermionic fields propagating in this manifold. Our main objective is to calculate the vacuum expectation values (VEV) of the  energy-momentum tensor, $\langle T^\mu_\nu\rangle$, associated with both matters fields. As will see, these VEVs are even function of  the magnetic flux running along the string's core, with period equal to the quantum flux $\Phi_0=\frac{2\pi}{e}$. Because the results obtained are expressed in terms of special functions, and a clear behavior for them are not very enlightening, we provide some graphs exhibiting their behavior as  function of the distance to the string considering different values for the parameters presented in the model considered.

\end{abstract}
\bigskip
%
\section{Introduction}
\label{Int}
%
\hspace{0.55cm}The physics underlying quantum vacuum fluctuations arises once quantum aspects of relativistic phenomena are taken into account. That means a quantized relativistic field will have a fluctuating ground state. In Minkowski spacetime, for instance, the renormalized vacuum expectation value (VEV) of physical observables, as a consequence of quantum vacuum fluctuations of relativistic fields presents a non-vanishing value if somehow the vacuum state is modified by some external influences. These external influences are in general boundary conditions obeyed by the quantum fields, or by coupling them with external interactions. One very known physical observable that gets a nonzero VEV due to the boundary conditions is the energy associated to the Casimir effect \cite{Mostepanenko:1997sw, bordag2009advances, Milton:2001yy}.

An additional feature related to the modifications of quantum vacuum fluctuations of relativistic fields is due to the presence of a curved background. It is well known that geometrical and topological aspects of a curved spacetime also induce a nonzero VEV of physical observables. 

De Sitter (dS) and anti de-Sitter (AdS) spacetimes, are curved spaces  solutions of the vacuum Einstein equations in the presence of cosmological constant, $\Lambda$. Due to the fact that they are maximally symmetric numerous physical relevant problem can be exactly solvable in these backgrounds\footnote{These spacetimes enjoy the same degree of symmetry as the Minkowski one\cite{Birrel_Davies}.}. Moreover the importance of dS space  increased by the appearance of the inflationary cosmology scenario \cite{Linde}. As to the AdS space it plays a crucial role in two exciting developments in theoretical physics of the past decade such as the AdS/CFT correspondence and the braneworld scenario with large extra dimensions. The AdS/CFT correspondence (for a review see \cite{Aharony}) represents a realization of the holographic principle and relates string theories or supergravity in the AdS bulk with a conformal field theory living on its boundary.

It is believed that the Universe has been underwent several symmetry breaking phase transitions during its expansion. In these transitions different types of topological objects may have been created, like global monopoles and cosmic strings \cite{Kibble,V-S}. In particular, cosmic strings are of special interest. Although recent observational data on the cosmic microwave background radiation have ruled out cosmic strings as the primary source for primordial density
perturbation, they are still candidate for the generation of a number of
interesting physical effects like gamma ray bursts \cite{Berezinski},
gravitational waves \cite{Damour} and high energy cosmic rays \cite{Bhattacharjee}. In addition cosmic string has also been considered in scenarios beyond the standard model of particle physics, like in supersymmetry and in string theory approaches \cite{Hind,CopelandJ}.  Phenomenologically, current observations of CMB suggest cosmic strings can contribute to a small percentage of the primordial density perturbations \cite{Ade:2013xla} in the Universe.  More recently, cosmic strings have attracted renewed interest due to a variant of their formation mechanism that is proposed in the framework of brane inflation \cite{Sarangi}-\cite{Dvali}.

The geometry of the spacetime produced by a cosmic string can be approximately described by a planar angle deficit on the two-dimensional sub-space perpendicular to it \cite{VS}. Although this object was first introduced in the literature as being created by a Dirac-delta type distribution of energy and axial stress  along a straight infinity line, it can also  be described by classical field theory where the energy-momentum tensor associated with the Maxwell-Higgs  system, investigated by Nielsen and Olesen in \cite{Nielsen197345}, couples to the Einstein's equations. This coupled system was first investigated in \cite{PhysRevD.32.1323} and \cite{Linet1987240}.

In this paper we will analyze the vacuum polarization effects associated with charged bosonic and fermionic quantum fields in the higher-dimensional AdS spacetime in the presence of an idealized cosmic string carrying  magnetic flux running along its core. As will see this specific combined geometry  makes possible to identify in both observables above mentioned, the contributions come from each parts, namely, from the AdS geometry itself and from the conical topology produced by the cosmic string.\footnote{The analysis of the vacuum polarization induced by an idealized cosmic string in dS spacetime, has bee developed in \cite{Saharian}} 

This paper is organized as follows. In section \ref{sec2} we analyze the effects of the quantum vacuum associated with a charged bosonic field in higher dimension AdS spacetime in the presence of a carrying magnetic flux cosmic string. Specifically in subsection \ref{Wightman_fun}  we present the geometry of the background spacetime that we want to work and construct the positive frequency Wightman function by the obtained complete set of normalized solutions of the  Klein-Gordon equation. We explicitly show that this Wightman function is decomposed in two contributions: the first induced by the AdS curved space, and the second contributions by the cosmic string. Having the positive frequency Wightman function in subsection \ref{Field_squared} we calculate the formal expression for the VEV of the field squared. Specifically our main objective is to analyze the contribution for this quantity induced by the cosmic string.  Some asymptotic behaviors of this observable are presented, and also a plot exhibiting its behavior as function of the distance from the string's core. The analysis of the VEV of the energy-momentum tensor induced by the cosmic string is developed in \ref{Energy-momentum}. In this subsection we show that our results obey the conservation condition. Also we present the expression for the energy-density for conformally coupled massless field. The behaviors of the energy-density as function of the distance to the string  considering different values of the physical parameters are presented in a separated graph. In section \ref{sec3} we investigate the quantum effects associated with a charged fermionic field in a $(1+4)-$ dimensional AdS space in the presence of a cosmic string. Using the Dirac equation defined in a curved space coupled with a four-vector potential, we obtain the normalized positive and negative energy solutions in subsection \ref{Dirac_eq}. In subsection \ref{condensate} we calculate the fermionic condensate by using the sum over the complete set of fermionic mode functions. Also in this subsection, we shown that this observable is decomposed in two contributions; the first one induced by the AdS space only, and the other is the contribution induced by the presence of the cosmic string. Here in this paper we are interested to analyze fermionic condensate induced by the cosmic string. Two plots exhibiting the behavior of this quantity as function of the distance to the string are provided.  Finally in subsection \ref{Fermionic_energy-momentum} we calculate the VEV of the energy-momentum tensor. Special attention is devoted to the cosmic string's induced part. We prove that this contribution satisfies the conservation condition  and the trace relation. In addition we present plots that exhibit the behavior of the energy-density as function the distance to the string, and also as function of the ration of magnetic flux by the quantum one and the product of the mass of fermion times the parameter that determines the curvature of the AdS space. In Section \ref{sec4} we summarize our most relevant results. In this paper we shall use the units $\hbar =G=c=1$.

\section{Quantum vacuum effects associated with charged bosonic fields}
\label{sec2}
This section is devoted to analyze the quantum fluctuations associated with a charged bosonic field in the higher-dimensional AdS curved space in the presence of a carrying-magnetic flux cosmic string. Specifically we  calculate the VEV of the field squared and the energy-momentum tensor. This is the plan of this section. However first it is necessary to obtain the complete set of  normalized wave-function. This calculation is presented in the next subsection.  
\subsection{Klein-Gordon equation and Wightman function}
\label{Wightman_fun}
The main objective of this subsection is to obtain the positive frequency Wightman function associated with a massive scalar charged field in a $(1+D)$-dimensional AdS spacetime, with $D\geq3$, considering the presence of a cosmic string. With this function we can calculate the vacuum polarization effects. In order to do that we first obtain the complete set of normalized bosonic modes for the Klein-Gordon equation admitting an arbitrary curvature coupling parameter, and the presence of an axial vector potential. 

The best coordinate system to represent the geometry under consideration is the cylindrical one. In a $(3+1)$-dimensional AdS spacetime in the presence of a cosmic string is given by the line element below:
\begin{equation}
ds^{2}=e^{-2y/a}[dt^{2}-dr^{2}-r^{2}d\phi ^{2}]-dy^{2}\ ,  \label{ds1}
\end{equation}
where $r\geqslant 0$ and $\phi \in \lbrack 0,\ 2\pi /q]$ define the
coordinates on the conical geometry, $(t, \ y)\in (-\infty ,\ \infty )$, and
the parameter $a$ determines the curvature scale of the AdS curved space. Here we consider a static string along the $y$-axis. The parameter $q$ is related to the mass per unit length $\mu $ of the string by the formula $q^{-1}=1-4G\mu $, where $G$ is the Newton's
gravitational constant. It encode the presence a planar angle deficit given by $\Delta\phi=2\pi(1-q^{-1})$ in this geometry. Using the \textit{Poincar\'{e}} coordinate \cite{OliveiradosSantos:2018spy,OliveiradosSantos:2019rjt} defined by $w=ae^{y/a}$, the \eqref{ds1} can be written in the form conformally related to the line element associated with a cosmic string in Minkowski spacetime, as shown below:
\begin{equation}
ds^2 = \left(\frac{a}{w}\right)^2[dt^2 - dr^2 - r^2d\phi^2 - dw^2 ] 
\label{ds2}
\end{equation}
The new coordinate  $w\in \lbrack 0,\ \infty )$. Two specific values for this coordinates should be mentioned: $w=0$ and $w=\infty $. They correspond to the AdS boundary and horizon, respectively.

The generalization of (\ref{ds2}) to $(D+1)$-dimensional AdS spacetimes is done
in the usual way, by adding extra Euclidean coordinates \cite{deMello:2011ji}:
\begin{equation}
\label{HDCS}
ds^2 = \left(\frac{a}{w}\right)^2\bigg[dt^2 - dr^2 - r^2d\phi^2 - dw^2 - \sum_{i=4}^{D}(dx^i)^2\bigg]   \   .
\end{equation}
   
 The curvature scale $a$ in \eqref{HDCS} is related to the cosmological constant, $\Lambda $, and the Ricci scalar, $R$, by the formulas
\begin{equation}
\Lambda =-\frac{D(D-1)}{2a^{2}} \ ,\ \ R=-\frac{D(D+1)}{a^{2}}\ .
\label{LamR}
\end{equation}

The field equation which governs the quantum dynamics of a charged
bosonic field with mass $m$, in a curved space and in the presence of an 
electromagnetic potential vector, $A_\mu$, is written by,
\begin{equation}
	(g^{\mu\nu}D_{\mu}D_{\nu} + m^2 + \xi R)\varphi(x) = 0  \   , 
	\label{KGE}
\end{equation}
where $D_{\mu}=\partial_{\mu}+ieA_{\mu}$. In the equation above we have considered the presence of a non-minimal coupling, $\xi$, between the field and the geometry represented by the Ricci scalar, $R$. Two specific values for the curvature coupling are $\xi = 0$ and $\xi = \frac{D - 1}{4D}$, that correspond to minimal and conformal coupling, respectively. As to the vector potential we have,
\begin{equation}
	A_{\mu} = A_{\phi}\delta^2_\mu   \  ,
	\label{VP}
\end{equation}
with $A_{\phi}=-q\Phi/(2\pi)$, being $\Phi$ the magnetic fluxes along the string's core. 

In the spacetime defined by \eqref{HDCS} and in the presence of the vector 
potentials given above, the equation \eqref{KGE} becomes,
\begin{eqnarray}
	\left[\frac{\partial^2}{\partial t^2} - \frac{\partial^2}{\partial r^2} - \frac{1}{r}\frac{\partial}{\partial r} - \frac{1}{r^2}\left(\frac{\partial}{\partial\phi} + ieA_{\phi}\right)^2 -	 \frac{\partial^2}{\partial w^2}\right.\nonumber\\
	\left.-\frac{(1-D)}{w}\frac{\partial}{\partial w}
 + \frac{M(D,m,\xi)}{w^2} - \sum_{i=4}^{D}\frac{\partial^2}{\partial (x^i)^2} \right]\varphi(x) = 0  \  . 
	\label{KGE2}
\end{eqnarray}
where $M(D,m,\xi) = a^2m^2 - \xi D(D+1)$. 

The positive energy solution of the equation above regular at origin can be obtained by factoring the wave-function in functions that depend on their respective coordinates. After some intermediate steps we obtain, 
\begin{equation}
	\varphi_\sigma(x) = Cw^{\frac{D}{2}}J_{\nu}(pw)J_{q|n +\alpha|}(\lambda r)e^{-iE t + iqn\phi + i\vec{k}\cdot\vec{x}_{\parallel}}.
	\label{Solu1}
\end{equation}
In the expression above $\vec{x}_{\parallel}$ represents the coordinates along the $(D-4)$ extra dimensions, and $\vec{k}$ the corresponding momentum. Moreover,
\begin{eqnarray}
	\nu &=& \sqrt{\frac{D^2}{4} + a^2m^2 - \xi D(D+1)},\nonumber\\
	E &=& \sqrt{\lambda^2 + p^2 + \vec{k}^2 },\nonumber\\
	\alpha &=& \frac{eA_{\phi}}{q} = -\frac{\Phi_{\phi}}{\Phi_0}.
	\label{const}
\end{eqnarray}
where $\Phi_0=\frac{2\pi}{e}$, the quantum flux. In \eqref{Solu1} $J_\mu(z)$ represents the Bessel function \cite{Abra}. The wave-function above is characterized by the set of quantum numbers, $\sigma=\{p, \ \lambda, \ {\vec{k}}, \ n\}$.

The constant $C$ in \eqref{Solu1} can be obtained by using the standard normalization condition for bosonic wave-function, 
\begin{eqnarray}
	\int d^Dx\sqrt{|g|}g^{00}\varphi_{\sigma'}^{*}(x)\varphi_{\sigma}(x)= \frac{1}{2E}\delta_{\sigma,\sigma'}  \   ,
	\label{NC}
\end{eqnarray}
where the delta symbol on the right-hand side is understood as Dirac delta 
function for the continuous quantum number, $\lambda$, $p$ and ${\vec{k}}$, and
Kronecker delta for the discrete one, $n$. From \eqref{NC} one finds 
\begin{eqnarray}
	|C|= \sqrt{\frac{qa^{1-D}\lambda p}{2E (2\pi)^{D-2}}}.
	\label{NC3}
\end{eqnarray}
Substituting the above normalization constant in the wave-function we obtain,
\begin{equation}
	\varphi_{\sigma}(x) =  \sqrt{\frac{qa^{1-D}\lambda p}{2E (2\pi)^{D-2}}}w^{\frac{D}{2}}J_{\nu}(pw)J_{q|n +\alpha|}(\lambda r)e^{-iE_{l} t + iqn\phi +i\vec{k}\cdot\vec{x}_{\parallel}}  \  .
	\label{COS}
\end{equation}

The properties of the vacuum state can be given by the positive 
frequency Wightman function, $W(x,x')=\left\langle 0|\hat{\varphi}(x) \hat{\varphi}^{*}(x')|0 \right\rangle$, where $|0 \rangle$ stands for the vacuum state. In order to evaluate  this function we adopt the mode sum below:
\begin{equation}
	W(x,x') = \sum_{\sigma}\varphi_{\sigma}(x)\varphi_{\sigma}^{*}(x'),
	\label{wight}
\end{equation}
where $\sum_{\sigma}$ represents integration over continuous quantum numbers, $p$, $\lambda$ and $\vec{k}$, and summation over $n$. 
Substituting \eqref{COS} into \eqref{wight} we obtain,
\begin{eqnarray}
	W(x,x') = \frac{qa^{1-D}(ww')^{\frac{D}{2}}}{2(2\pi)^{D-2}}\sum_{n=-\infty}^{\infty}e^{inq\Delta\phi}\int d\vec{k}\int_0^{\infty}dpp\int_0^{\infty}d\lambda\lambda\nonumber\\
	\times J_{q|n +\alpha|}(\lambda r)J_{q|n +\alpha|}(\lambda r')J_{\nu}(pw)J_{\nu}(pw')\frac{e^{-iE\Delta t+ i\vec{k}\cdot\Delta\vec{x}_{\parallel}}}{E}
	\label{wight2}
\end{eqnarray}
where $\Delta t=t-t', \Delta \phi=\phi-\phi'$ and $\Delta \vec{x}_{\parallel}= \vec{x}_{\parallel}-\vec{x}_{\parallel}'$.

Now performing a Wick rotation, and using the identity
\begin{equation}
	\frac{e^{-\Delta\tau\omega}}{\omega}=\frac2{\sqrt{\pi}}\int_0^\infty ds e^{-s^2\omega^2-\Delta\tau^2/(4s^2)}  \  ,
	\label{identity}
\end{equation} 
we rewrite the Wightman function as,
\begin{eqnarray}
	W(x,x')&=&\frac{q(ww')^{\frac{D}{2}}}{(2\pi)^{D-2}a^{D-1}}\int d\vec{k}e^{i\vec{k}\cdot\Delta\vec{x}_{\parallel}}\int_0^{\infty}dppJ_{\nu}(pw)J_{\nu}(pw')\nonumber\\
	&\times&\sum_{n=-\infty}^{\infty}e^{inq\Delta\phi}\int_0^{\infty}d\lambda\lambda J_{q|n +\alpha|}(\lambda r)J_{q|n +\alpha|}(\lambda r')\nonumber\\&\times&\int_{0}^{\infty}\frac{ds}{s}e^{-s^2(\lambda^2+p^2+\vec{k}^2)-(\Delta z^2-\Delta t^2)/4s^2} \  .
	\label{propagator-to-sum}
\end{eqnarray}
The integral over $\vec{k}$ in the expression above is trivial. Using the results given in \cite{Grad},
\begin{equation}
	\int_0^{\infty}d\eta\eta e^{-\eta^2s^2}J_{\gamma}(\eta\rho)J_{\gamma}(\eta\rho') = \frac{e^{-\frac{(\rho^2 + \rho'^2)}{4s^2}}}{2s^2}I_{\gamma}\left(\frac{\rho\rho'}{2s^2}\right)  \  ,
	\label{id2}
\end{equation}
we can integrate over $\lambda$ and $p$, obtaining
\begin{eqnarray}
	W(x,x')&=&\frac{q a^{1-D}}{2(2\pi)^{\frac{D}{2}}}\bigg(\frac{ww'}{rr'}\bigg)^{\frac{D}{2}}\int_{0}^{\infty}d\chi \chi^{\frac{D}{2}-1}e^{-\chi u^{2}/2rr'}
	I_{\nu}\bigg(\frac{ww'}{rr'}\chi\bigg)\nonumber\\
	&\times&\sum_{n=-\infty}^{\infty}e^{iqn\Delta\phi}I_{q|n+\alpha|}(\chi)  \  , 
	\label{propagator-to-sum_a}
\end{eqnarray}
where we have introduced a new variable $\chi=rr'/2s^2$ and defined
\begin{equation}
	u^{2}=r^2+r'^2+w^2+w'^2+\Delta \vec{x}^{2}_{\parallel}-\Delta t^2  \ .
\end{equation}
The parameter $\alpha$ in Eq.\eqref{const} can be written in the form
\begin{equation}
	\alpha=n_{0}+\varepsilon, \ \textrm{with}\ |\varepsilon|<\frac{1}{2},
	\label{const-2}
\end{equation}
being $n_{0}$ an integer number. Using the result obtained in \cite{deMello:2014ksa}, we can develop the sum  over the quantum number $n$:
\begin{eqnarray}
	&&\sum_{n=-\infty}^{\infty}e^{iqn\Delta\phi}I_{q|n+\alpha|}(\chi)=\frac{1}{q}\sum_{k}e^{\chi\cos(2\pi k/q-\Delta\phi)}e^{i\alpha(2\pi k -q\Delta\phi)}\nonumber\\
	&-&\frac{e^{-iqn_{0}\Delta\phi}}{2\pi i}\sum_{j=\pm1}je^{ji\pi q|\varepsilon|}
	\int_{0}^{\infty}dy\frac{\cosh{[qy(1-|\varepsilon|)]}-\cosh{(|\varepsilon| qy)e^{-iq(\Delta\phi+j\pi)}}}{e^{\chi\cosh{(y)}}\big[\cosh{(qy)}-\cos{(q(\Delta\phi+j\pi))}\big]},
	\label{summation-formula}
\end{eqnarray}
where
\begin{equation}
	-\frac{q}{2}+\frac{\Delta\phi}{\Phi_{0}}\le k\le \frac{q}{2}+\frac{\Delta\phi}{\Phi_{0}}  \   .
\end{equation}
Substituting \eqref{summation-formula} into \eqref{propagator-to-sum_a}, and develop the integration over $\chi$ with the help of \cite{Grad}, we get,
\begin{eqnarray}
	W(x,x')&=&\frac{a^{1-D}}{(2\pi)^{\frac{D+1}{2}}}\Bigg\{\sum_{k}e^{i\alpha(2\pi k-q\Delta\phi)}F_{\nu-1/2}^{(D-1)/2}({u}_{k})-q\frac{e^{-iqn_{0}\Delta\phi}}{2\pi i}\nonumber\\
	&\times&\sum_{j=\pm1}je^{ji\pi q|\varepsilon|}
	\int_{0}^{\infty}dy\frac{\cosh{[(1-|\varepsilon|)qy]}-\cosh{(|\varepsilon|q y)e^{-iq(\Delta\phi+j\pi)}}}{\cosh{(qy)}-\cos{(q(\Delta\phi+j\pi))}}\nonumber\\
	&\times&F_{\nu-1/2}^{(D-1)/2}({u}_{y})\Bigg\}  \  .
	\label{full-propagator}
\end{eqnarray}
In the expression above we have introduced the notation
\begin{eqnarray}
	F_{\gamma}^{\mu}(u)&=&e^{-i\pi\mu}\frac{Q^{\mu}_{\gamma}(u)}{(u^2-1)^{\mu/2}}  \nonumber\\  
	&=&\frac{\sqrt{\pi}\Gamma(\gamma+\mu+1)}{2^{\gamma+1}\Gamma(\gamma+3/2)u^{\gamma+\mu+1}}F\bigg(\frac{\gamma+\mu}{2}+1,\frac{\gamma+\mu+1}{2};\gamma+\frac{3}{2};\frac{1}{u^{2}}\bigg).
	\label{function-2}
\end{eqnarray}
being $Q_{\gamma}^{\mu}(u)$ the associated Legendre function of second kind and $F(a,b;c;z)$ the hypergeometric function \cite{Abra}. In \eqref{full-propagator}, the arguments of the function $F_{\gamma}^{\mu}$ are given by
\begin{eqnarray}
	u_{k}&=&1+\frac{r^2+r'^2-2rr'\cos{(2\pi k/q-\Delta\phi)}+\Delta w^2+\Delta\vec{x}^{2}_{\parallel}-\Delta t^2}{2ww'}\nonumber\\
	u_{y}&=&1+\frac{r^2+r'^2+2rr'\cosh{(y)}+\Delta w^2+\Delta\vec{x}^{2}_{\parallel}-\Delta t^2}{2ww'}.
\end{eqnarray}
The Wightman function can be written as a sum of two components,
\begin{equation}
	W(x,x')=W_{\rm{AdS}}(x,x')+W_{\rm{cs}}(x,x') \  ,
	\label{wightman-function-expanded}
\end{equation}
where the first term, represented by $k=0$ term of the sum is the pure AdS spacetime contribution, and the second one, $k\neq0$, together with the integral is due to the presence of the cosmic string.
\subsection{The VEV of the Field Squared}
\label{Field_squared}
The VEV of the field squared is formally obtained by the evaluation of the Wightman function at coincidence limit, as shown below:
\begin{equation}
	\langle|\varphi|^2\rangle=\lim\limits_{x'\rightarrow x}W(x,x') \  .
\end{equation}
By substituting \eqref{wightman-function-expanded} into the above expression, we see that $\langle|\varphi|^2\rangle$ presents three contributions: 
\begin{equation}
	\langle|\varphi|^2\rangle=\langle|\varphi|^2\rangle_{\rm{AdS}}+\langle|\varphi|^2\rangle_{\rm{cs}} \  .
\end{equation}
Unfortunately the above expression provides a divergent result and needed to be renormalized. Due to the fact that the presence of a cosmic string does not introduce additional curvature for points outside of the string's core, the divergence comes only from the contribution due to the pure AdS spacetime. The analysis of the renormalized VEV of the field squared in AdS space has been developed in the literature \cite{Burgess}-\cite{Caldarelli}. So, in this paper we will analyze only the contribution induced by the string. 

Considering only the cosmic string component of the Wightman function and taking its  coincidence limit, we have 
\begin{eqnarray}
	\langle|\varphi|^2\rangle_{\rm{cs}}&=&\frac{2}{(2\pi)^{\frac{D+1}{2}}a^{D-1}}\Bigg[\sideset{}{'}\sum_{k=1}^{[q/2]}\cos{(2\pi k\varepsilon)}F^{(D-1)/2}_{\nu-1/2}(v_{k})\nonumber\\
	&-&\frac{q}{\pi}\int_{0}^{\infty}dy\frac{f(q,\varepsilon,2y)}{\cosh(2qy)-\cos(q\pi)}F^{(D-1)/2}_{\nu-1/2}(v_{y})\Bigg]  \  ,
	\label{phi-squared}
\end{eqnarray}
where we have introduced new notations:
\begin{eqnarray}
	v_{k}&=&1+2(r/w)^2\sin^2{(\pi k/q)}, \nonumber\\
	v_{y}&=&1+2(r/w)^2\cosh^2{(y)}  \  ,
\end{eqnarray}
being, 
\begin{equation}
	f(q,\varepsilon,2y)=\sin(|\varepsilon|q\pi)\cosh((1-|\varepsilon|)2qy)+\cosh(2|\varepsilon|qy)\sin((1-|\varepsilon|)q\pi) \ .
	\label{ffunction}
\end{equation}
In Eq.\eqref{phi-squared} $[q/2]$ represents the integer part of $q/2$ and the prime on the sign of the summation means that in the case $q=2p$ the term $1/2$ should be taken with the coefficient $1/2$.

From \eqref{phi-squared} it is possible to obtain some interesting asymptotic behaviors. Let us start taking $r/w\rightarrow0$. We can use the asymptotic expression for the hypergeometric function for small arguments \cite{Abra} to rewrite Eq.\eqref{phi-squared} as
\begin{equation}
	\langle|\varphi|^2\rangle_{cs}\approx\frac{2\Gamma\big(\frac{D-1}{2}\big)}{(4\pi)^{\frac{D+1}{2}}}\bigg(\frac{w}{ar}\bigg)^{D-1}\bigg[\sideset{}{'}\sum_{k=1}^{[q/2]}\frac{\cos(2\pi k\varepsilon)}{\sin^{D-1}(\pi k/q)}-\frac{q}{\pi}\int_{0}^{\infty}dy\frac{f(q,\varepsilon,2y)\cosh^{1-D}(y)}{\cosh(2qy)-\cos(q\pi)}\bigg] \ .
	\label{field-squared-cs-assymp}
\end{equation}
We can notice that the above result diverges with inverse of proper distance from the string at the power $(D-1)$. In other words, for a fixed value of $w$, $\langle|\varphi|^2\rangle_{cs}$ goes to infinity  for points near the string.

In the opposite limit, $r/w\to\infty$, by using the asymptotic expression below \cite{deMello:2014hya},
\begin{equation}
	F_{\nu-1/2}^{(D-1)/2}(u)\approx\frac{\sqrt{\pi}\Gamma(D/2+\nu)}{2^{\nu+1/2}\Gamma(\nu+1)u^{D/2+\nu}} \ ,
\end{equation}
we get
\begin{eqnarray}
	\langle|\varphi|^2\rangle_{cs}&\approx&\frac{2^{-2\nu}\Gamma(D/2+\nu)}{(4\pi)^{\frac{D}{2}}\Gamma(\nu+1)a^{D-1}}\bigg(\frac{w}{r}\bigg)^{D+2\nu}\bigg[\sideset{}{'}\sum_{k=1}^{[q/2]}\frac{\cos(2\pi k\varepsilon)}{\sin^{D+2\nu}(\pi k/q)}\nonumber\\
	&-&\frac{q}{\pi}\int_{0}^{\infty}dy\frac{f(q,\varepsilon,2y)\cosh^{-D-2\nu}(y)}{\cosh(2qy)-\cos(q\pi)}\bigg] \ .
	\label{field-squared-cs-assymp2}
\end{eqnarray}
From the above expression, we observe that for fixed values of the radial coordinate $r$, the string-induced contribution goes to zero near the AdS boundary as $w^{D+2\nu}$.

For a conformally coupled massless scalar field we have $\nu=1/2$. So, by expressing the associated Legendre function in terms of  hypergeometric function \cite{Abra,Grad}, we can  write a more convenient expression for $F_{\nu-1/2}^{(D-1)/2}(u)$ \cite{deMello:2014hya}, given by
\begin{eqnarray}
	F_{0}^{(D-1)/2}(u)=-\frac{\Gamma\big(\frac{D-1}{2}\big)}{2}\bigg[(1+u)^{-(D-1)/2}-(u-1)^{-(D-1)/2}\bigg].
	\label{function-3}
\end{eqnarray} 
Substituting \eqref{function-3} into \eqref{phi-squared}, we obtain
\begin{equation}
	\langle|\varphi|^2\rangle_{\rm{cs}}=\bigg(\frac{w}{a}\bigg)^{D-1}\bigg[\langle|\phi|^2\rangle_{\rm{cs}}^{\rm{(M)}}+\langle|\phi|^2\rangle_{\rm{cs,b}}^{\rm{(M)}}\bigg],
	\label{conformal-phi-squared}
\end{equation}
where
\begin{equation}
	\langle|\varphi|^2\rangle_{\rm{cs}}^{\rm{(M)}}=\frac{2\Gamma(\frac{D-1}{2})}{(4\pi)^{\frac{D+1}{2}}r^{D-1}}\Bigg[\sideset{}{'}\sum_{k=1}^{[q/2]}\frac{\cos{(2\pi k\varepsilon)}}{\sin^{D-1}(\pi k/q)}-\frac{q}{\pi}\int_{0}^{\infty}dy\frac{f(q,\varepsilon,2y)\cosh^{1-D}(y)}{\cosh(2qy)-\cos(q\pi)}\Bigg]
	\label{conformal-sf-cs}
\end{equation}
is the VEV for the boundary-free cosmic string geometry corrected by the presence of a magnetic flux running through the string's core.
The second term in \eqref{conformal-phi-squared},
\begin{eqnarray}
	\langle|\varphi|^2\rangle_{\rm{cs,b}}^{\rm{(M)}}&=&-\frac{2\Gamma(\frac{D-1}{2})}{(4\pi)^{\frac{D+1}{2}}}\Bigg[\sideset{}{'}\sum_{k=1}^{[q/2]}\cos(2\pi k\varepsilon)\bigg(w^2+r^2\sin^2(\pi k/q)\bigg)^{-\frac{(D-1)}{2}}\nonumber\\
	&-&
	\frac{q}{\pi}\int_{0}^{\infty}dy\frac{f(q,\varepsilon,2y)}{\cosh(2qy)-\cos(q\pi)}\bigg(w^2+r^2\cosh^2(y)\bigg)^{-\frac{(D-1)}{2}}\Bigg],
\end{eqnarray}
is the contribution induced by the boundary located at $w=0$. It is finite at the string's core for $w\neq 0$. In addition, for $r\gg w$ this contribution tends to cancel \eqref{conformal-sf-cs} inside the square bracket in Eq.\eqref{conformal-phi-squared}.

\begin{figure}[!htb]
	\begin{center}
		\centering
		\includegraphics[scale=0.4]{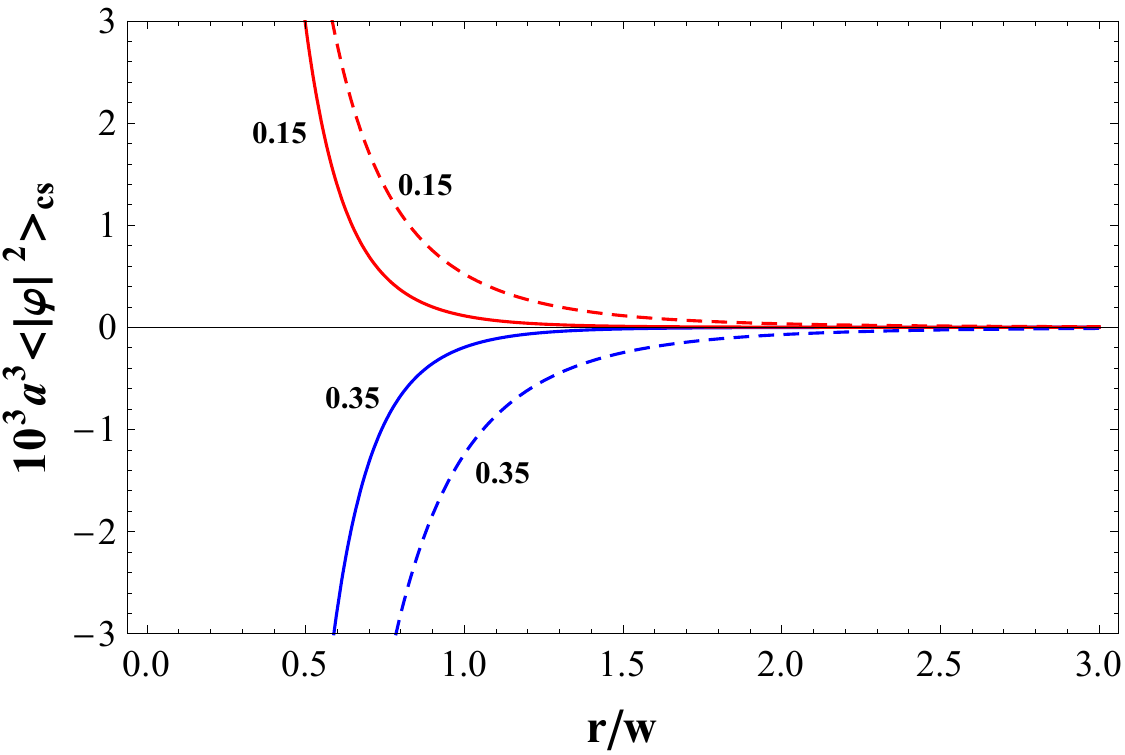}
		\caption{The VEV of the field squared without compactification for $D=4$ in Eq.\eqref{phi-squared} is plotted in units of $a^{3}$, in terms of $r/w$, for $q=2.5$ and $m=0$. The numbers near the curves correspond to the values of the parameter $\varepsilon$. The full lines correspond to the massless minimum coupling field while the dashed ones correspond to the conformal massless case.}
		\label{fig1_0}
	\end{center}
\end{figure}

In Fig.\ref{fig1_0} we present the behavior of $\langle|\varphi|^2\rangle_{cs}$  as a function of $r/w$, that is the proper distance from the string in units of AdS curvature radius. By this plot, we note how the intensity and and behavior of the field squared change with the parameter $\varepsilon$ and $\xi$. 

\subsection{The VEV of the Energy-Momentum tensor}
\label{Energy-momentum}
Our next step is to obtain the VEV of the energy-momentum tensor. We can develop this calculation by using the Wightman function and the VEV of the field squared. In order to do that we will use the the expression obtained in \cite{OliveiradosSantos:2019rjt}:
\begin{equation}
	\langle T_{\mu\nu}\rangle=(D_{\mu}D_{\nu'}^{\dagger}+D_{\mu'}^{\dagger}D_{\nu})W(x,x')-2[\xi R_{\mu\nu}+\xi\nabla_{\mu}\nabla_{\nu}-(\xi-1/4)g_{\mu\nu}\nabla_{\alpha}\nabla^{\alpha}]]\langle|\varphi|^2\rangle,
	\label{E-M-Tensor-formula}
\end{equation}
where $R_{\mu\nu}=-Dg_{\mu\nu}/a^2$ is the Ricci tensor for the AdS spacetime and $D_{\mu}=\nabla_{\mu}+ieA_{\mu}$. Similarly to the VEV of the field squared, the VEV of the energy-momentum tensor can be decomposed as
\begin{equation}
	\langle T_{\mu\nu}\rangle=\langle T_{\mu\nu}\rangle_{\rm{AdS}}+\langle T_{\mu\nu}\rangle_{\rm{cs}} \  .
	\label{TEM0}
\end{equation}
The first contribution has been analyzed in \cite{Caldarelli}. Here in this paper, we are  interested in the calculation of the contribution due to the string only.

The covariant d'Alembertian acting in \eqref{phi-squared} provides,
\begin{eqnarray}
	\Box\langle |\varphi|^2\rangle_{\rm{cs}}&=&-\frac{8}{(2\pi)^{\frac{D+1}{2}}a^{D+1}}\Bigg[\sideset{}{'}\sum_{k=1}^{[q/2]}\cos(2\pi k\varepsilon)g(v_{k},\sin(\pi k/q))\\\nonumber
	&-&\frac{q}{\pi}\int_{0}^{\infty}dy\frac{f(q,\varepsilon,2y)g(v_{y},\cosh(y))}{\cosh(2qy)-\cos(q\pi)}\Bigg]
\end{eqnarray}
where the function $g$ is defined below,
with the notation
\begin{equation}
	g(u,v)=(u-1)(2v^2+u-1)\frac{d^2}{du^2}F^{(D-1)/2}_{\nu-1/2}(u)+\bigg[2v^2+\frac{D+2}{2}(u-1)\bigg]\frac{d}{du} F^{(D-1)/2}_{\nu-1/2}(u).
\end{equation}

For the geometry under consideration, only the differential operators, $\nabla_{r}\nabla_{w}$ and $\nabla_{\mu}\nabla_{\mu}$, provide a non vanishing contribution when acting on the VEV of the field squared. The remaining contributions to the energy-momentum tensor come from the electromagnetic covariant derivatives acting on the Wightman function. To analyze the azimuthal term, it is more convenient to act the operator $D_{\phi}D_{\phi'}^{\dagger}$ in \eqref{propagator-to-sum}, and them take the coincidence limit in the angular variable. Following this procedure, we obtain the following intermediate expression,
\begin{equation}
	S(q,\alpha,\chi)=\sum_{n=-\infty}^{\infty}q^2(n+\alpha)^2I_{q|n+\alpha|}(\chi),
\end{equation}
being $\chi=rr'/2s^2$. The above sum can be developed by using the differential equation obeyed by the modified Bessel equation.  So we can write,
\begin{equation}
	S(q,\alpha,\chi)=\bigg(\chi^2\frac{d^2}{d\chi^2}+\chi\frac{d}{d\chi}-\chi^2\bigg)\sum_{n=-\infty}^{\infty}I_{q|n+\alpha|}(\chi),
\end{equation}
where this last sum can be obtained from previous result \eqref{summation-formula}:
\begin{equation}
	\sum_{n=-\infty}^{\infty}I_{q|n+\alpha|}(\chi)=\frac{2}{q}\sideset{}{'}\sum_{k=0}^{[q/2]}\cos(2\pi k\alpha)e^{\chi\cos(2\pi k/q)}-\frac{2}{\pi}\int_{0}^{\infty}dy\frac{e^{-\chi\cosh(2y)}f(q,\varepsilon,2y)}{\cosh(2qy)-\cos(q\pi)}.
\end{equation}

The cosmic string contribution in the VEV of the energy-momentum tensor is calculated from previous expression, \eqref{E-M-Tensor-formula}, by making use of the string component of the Wightman function, $W_{cs}(x,x')$, and VEV of the field squared given in \eqref{phi-squared}. After long but straightforward calculations, for the string part, one finds (no summation over $\mu$)
\begin{eqnarray}
	\langle T_{\mu}^{\mu}\rangle_{\rm{cs}}&=&-\frac{4}{(2\pi)^{\frac{D+1}{2}}a^{D+1}}\Bigg[\sideset{}{'}\sum_{k=1}^{[q/2]}\cos(2\pi k\varepsilon)g_{\mu}^{(0)}(v_{0k},\sin(\pi k/q))\nonumber\\
	&-&\frac{q}{\pi}\int_{0}^{\infty}dy\frac{f(q,\varepsilon,2y)g_{\mu}^{(0)}(v_{0y},\cosh(y))}{\cosh(2qy)-\cos(q\pi)}\Bigg],
	\label{E-M-cosmic-string}
\end{eqnarray}
where
\begin{equation}
	g_{\mu}^{(0)}(u,v)=G_{\mu,0}^{\mu}(u,v)+(4\xi-1)g(u,v)-\xi DF^{(D-1)/2}_{\nu-1/2}(u),
\end{equation}
and
\begin{eqnarray}
	G_{0,0}^{0}(u,v)&=&-[1+2\xi (u-1)]\frac{d}{du}F^{(D-1)/2}_{\nu-1/2}(u)\nonumber\\
	G_{1,0}^{1}(u,v)&=&[2v^2(1-2\xi)-1-2\xi (u-1)]\frac{d}{du}F^{(D-1)/2}_{\nu-1/2}(u)\nonumber\\&+&2v^2(1-4\xi)(u-1)\frac{d^2}{du^2}F^{(D-1)/2}_{\nu-1/2}(u)\nonumber\\
	G_{2,0}^{2}(u,v)&=&-[1+2v^2(2\xi-1)+2\xi(u-1)]\frac{d}{du}F^{(D-1)/2}_{\nu-1/2}(u)\nonumber\\&-&2(u-1)(1-v^2)\frac{d^2}{du^2}F^{(D-1)/2}_{\nu-1/2}(u)\nonumber\\
	G_{3,0}^{3}(u,v)&=&[(1-4\xi)(u-1)-1]\frac{d}{du}F^{(D-1)/2}_{\nu-1/2}(u)\nonumber\\&+&(u-1)^2(1-4\xi)\frac{d^2}{du^2}F^{(D-1)/2}_{\nu-1/2}(u).
\end{eqnarray}
As to the other components with $\mu=4,...,D$, we have (no summation) $\langle T_{\mu}^{\mu}\rangle_{\rm{cs}}=\langle T_{0}^{0}\rangle_{\rm{cs}}$. 

For the non-zero off-diagonal component, we have
\begin{eqnarray}
	\langle T_{3}^{1}\rangle_{\rm{cs}}=-\frac{4}{(2\pi)^{\frac{D+1}{2}}a^{D+1}}\frac{w}{r}\Bigg[\sideset{}{'}\sum_{k=1}^{[q/2]}\cos(2\pi k\varepsilon)h^{(0)}(v_{k})-\frac{q}{\pi}\int_{0}^{\infty}dy\frac{f(q,\varepsilon,2y)h^{(0)}(v_{y})}{\cosh(2qy)-\cos(q\pi)}\Bigg],
	\label{E-M-cs-off-diagonal}
\end{eqnarray}
where
\begin{equation}
	h^{(0)}(u)=(u-1)\bigg[(1-2\xi)\frac{d}{du}F^{(D-1)/2}_{\nu-1/2}(u)+(u-1)(1-4\xi)\frac{d^2}{du^2}F^{(D-1)/2}_{\nu-1/2}(u)\bigg].
\end{equation}

It is possible to check that the energy-momentum tensor, $\langle T_{\mu}^{\nu}\rangle_{\rm{cs}}$, given above satisfies the trace identity:
\begin{equation}
	\langle T_{\mu}^{\mu}\rangle_{\rm{cs}}=2[D(\xi-\xi_{c})\nabla_{\mu}\nabla^{\mu}\langle|\varphi|^2\rangle_{\rm{cs}}+m^2\langle|\varphi|^2\rangle_{\rm{cs}}] \  .
	\label{EM-trace}
\end{equation}
Note it is traceless for a conformal massless quantum scalar field.
Furthermore, as an additional verification for the expressions in \eqref{E-M-cosmic-string} and off-diagonal component in \eqref{E-M-cs-off-diagonal}, we can see that the covariant conservation equation $ \nabla_{\mu}\langle T_{\nu}^{\mu}\rangle=0$ is obeyed. For the geometry under consideration the latter is expressed by the relations
\begin{equation}
	\frac{1}{r}\partial_{r}(r\langle T_{1}^{1}\rangle_{\rm{cs}})-\frac{1}{r}\langle T_{2}^{2}\rangle_{\rm{cs}}-\frac{D+1}{w}\langle T_{1}^{3}\rangle_{\rm{cs}}+\partial_{w}\langle T_{1}^{3}\rangle_{\rm{cs}}=0
	\label{c-eq-1}
\end{equation}
and
\begin{equation}
	\frac{1}{r}\partial_{r}(r\langle T_{3}^{1}\rangle_{\rm{cs}})+\partial_{w}\langle T_{3}^{3}\rangle_{\rm{c}}-\frac{D+1}{w}\langle T_{3}^{3}\rangle_{\rm{cs}}+\frac{1}{w}\langle T_{\mu}^{\mu}\rangle_{\rm{cs}}=0.
	\label{c-eq-2}
\end{equation}

For a conformally coupled massless scalar field, the energy density component, $\langle T_{0}^{0}\rangle_{\rm{cs}}$, reads
\begin{eqnarray}
	\langle T_{0}^{0}\rangle_{\rm{cs}}&=&\frac{2\Gamma(\frac{D+1}{2})}{(4\pi)^{\frac{D+1}{2}}Da^{D+1}}\Bigg\{\sideset{}{'}\sum_{k=1}^{[q/2]}\cos(2\pi k\varepsilon)\Bigg[Df^{\frac{{D-1}}{2}}(\rho_{k})\nonumber\\&-&\Bigg(D+2s_{k}^2+(2D+1)\rho_{k}\Bigg)f^{\frac{D+1}{2}}(\rho_{k})+(D+1)\Bigg(\frac{r^2s_{k}^4}{w^2}+\rho_{k}^2\Bigg)f^{\frac{D+3}{2}}(\rho_{k})\Bigg]\nonumber\\&-&\frac{q}{\pi}\int_{0}^{\infty}dy\frac{f(q,\varepsilon,2y)}{\cosh(2qy)-\cos(q\pi)}\Bigg[Df^{\frac{{D-1}}{2}}(\rho_{y})-\Bigg(D+2s_{y}^2+(2D+1)\rho_{y}\Bigg)\nonumber\\&\times&f^{\frac{D+1}{2}}(\rho_{y})+(D+1)\Bigg(\frac{r^2s_{y}^4}{w^2}+\rho_{y}^2\Bigg)f^{\frac{D+3}{2}}(\rho_{y})\Bigg]\Bigg\} \ ,
	\label{EMT-cs-comp00-conformal}
\end{eqnarray}
where we have introduced the function
\begin{equation}
	f^{\alpha}(\rho)=\frac{1}{(\rho_\gamma)^{\alpha}}-\frac{1}{(1+\rho_\gamma)^{\alpha}}
	\label{ffunction2}
\end{equation}
being
\begin{eqnarray}
	\rho_{\gamma}&=&\bigg(\frac{rs_{\gamma}}{w}\bigg)^2   \ ,
	\label{rho-var}
\end{eqnarray}
with $\gamma=s_{k}=\sin(k\pi/q)$ or $\gamma=s_{y}=\cosh(y)$.

As to the off-diagonal component  of the energy-momentum tensor, for the same situation as above, $\langle T_{3}^{1}\rangle_{\rm{cs}}$ reads,
\begin{eqnarray}
	\langle T_{3}^{1}\rangle_{\rm{cs}}&=&-\frac{4\Gamma(\frac{D+3}{2})}{(4\pi)^{\frac{D+1}{2}}Da^{D+1}}\frac{r}{w}\Bigg\{\sideset{}{'}\sum_{k=1}^{[q/2]}\cos(2\pi k\varepsilon)\sin^2(\pi k/q)\nonumber\\&\times&\bigg[1+\frac{r^2}{w^2}\sin^{2}(\pi k/q)\bigg]^{-\frac{D+3}{2}}-\frac{q}{\pi}\int_{0}^{\infty}dy\frac{f(q,\varepsilon,2y)\cosh^2(y)}{\cosh(2qy)-\cos(q\pi)}\nonumber\\&\times&\bigg[1+\frac{r^2}{w^2}\cosh^{2}(y)\bigg]^{-\frac{D+3}{2}}\Bigg\} \ .
	\label{conformal-off-diagonal-cs}
\end{eqnarray}
As you can see in the conformal coupled massless scalar field, the above equations are expressed in terms of elementary functions.

\begin{figure}[!htb]
	\begin{center}
		\centering
		\includegraphics[scale=0.5]{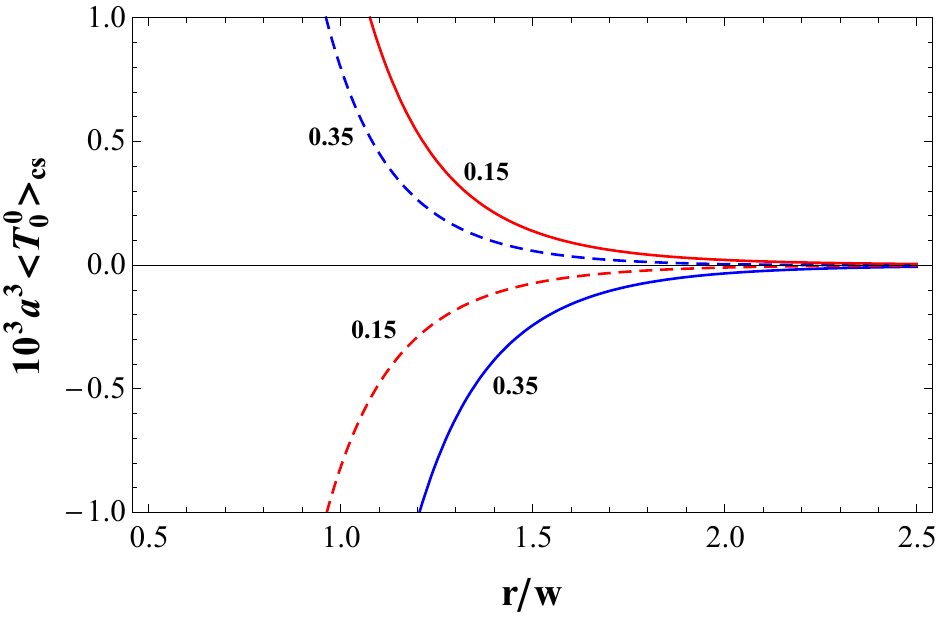}
		\caption{The VEV of the energy density in uncompactified string geometry is plotted for $D=4$, in units of $a^{3}$, in terms of $r/w$,  for $q=2.5$ and $m=0$. The numbers near the curves refer to the values of the parameter $\varepsilon$. The full lines correspond to the massless minimum coupling field while the dashed ones correspond to the conformal massless case.}
		\label{fig4}
	\end{center}
\end{figure}

In Fig. \ref{fig4} we have exhibited the behavior of the energy density induced by the presence of a  cosmic string geometry, $\langle T_{0}^{0}\rangle_{\rm{cs}}$, as function of  the proper distance from the string measured in units of the AdS curvature radius $a$. By this plot we can see that the parameters associated with magnetic flux along the string, $\varepsilon$, and the curvature coupling, $\xi$, can modify the intensity and behavior of the energy density.

\section{Quantum vacuum effects associated with charged fermionic fields}
\label{sec3} 
Here we will analyze the vacuum fluctuation associated with charged fermionic field in a $(1+4)-$dimensional AdS background in the presence of a cosmic string. Specifically we want to calculate the fermionic condensate and the VEV of the energy-momentum tensor.

\subsection{Dirac equation and fermionic modes}
\label{Dirac_eq}
The dynamics of a charged fermionic quantum field in curved space coupled with a gauge potential, $A_{\mu}$, is given by the following equation:
\begin{equation}
	\left( i\gamma ^{\mu }\mathcal{D}_{\mu }-sm\right) \psi (x)=0\  , \ {\rm with}  \ \ \mathcal{D}
	_{\mu }=\partial _{\mu }+\Gamma _{\mu }+ieA_{\mu } \ , \ s=\pm1 \ ,  \label{DiracEq}
\end{equation}
In the above equation $\gamma^{\mu}$ represent the Dirac matrices in curved space and $\Gamma_{\mu}$ the spin connection. Because want to work in $(1+4)-$dimensional spacetime, the two possibles values of $s$ correspond to the two irreducible representations of Dirac matrices. Both sets of matrices are related to the flat ones, $\gamma^{(a)}$, by the relations,
\begin{equation}
	\gamma_{\mu}=e^{\nu}_{(a)}\gamma^{(a)},\quad \Gamma_{\mu}=\frac{1}{4}\gamma^{(a)}\gamma^{(b)}e^{\nu}_{(a)}e_{(b)\nu,\mu}  \  .
\end{equation}
With the tetrad basis, $e^{\mu}_{(a)}$, satisfying the relation $e^{\mu}_{(a)}e^{\nu}_{(b)}\eta^{ab}=g^{\mu\nu}$,  being $\eta^{ab}$ the Minkowski spacetime metric tensor.

Now we want to introduce the geometry of the spacetime that we want to consider. Using again the \textit{Poincar\'{e}} the metric tensor can be obtained by the line element below,
\begin{equation}
	\label{metric}
	ds^2 = \left(\frac{a}{w}\right)^2\bigg[dt^2 - dr^2 - r^2d\phi^2 - dw^2 - dz^2\bigg]   \   .
\end{equation}

The set of Dirac matrices in flat spacetime assumed here is the following:
\begin{equation}
	\label{gamma_matrices_flat}
	\gamma^{(0)}=-i
	\left( {\begin{array}{cc}
			0 & 1 \\
			-1 & 0 \\
	\end{array} } \right), \quad \gamma^{(a)}=-i
	\left( {\begin{array}{cc}
			\sigma_{a} & 0 \\
			0 & -\sigma_{a} \\
	\end{array} } \right), \ \text{with} \  (a)=1,2,3, \quad \gamma^{(4)}=i\left( {\begin{array}{cc}
			0 & 1 \\
			1 & 0 \\
	\end{array} } \right),
\end{equation}
where $\sigma_{1},\sigma_{2},\sigma_{3}$ correspond to the Pauli matrices. Furthermore we take the tetrad basis as follows:
\begin{equation}
	e^{\mu}_{(a)}=\frac{w}{a}\left( {\begin{array}{ccccc}
			1 & 0 & 0 & 0 & 0\\
			0 & \cos(q\phi) & -\sin(q\phi)/r & 0 & 0\\
			0 & \sin(q\phi) & \cos(q\phi)/r & 0 & 0\\
			0 & 0 & 0 & 1 & 0\\
			0 & 0 & 0 & 0 & 1\\
	\end{array} } \right)
\end{equation}
where the index $(a)$ identifies the rows  of the matrix.

With this choice, the gamma matrices in the curved space take the form
\begin{equation}
	\gamma^{0}=\frac{w}{a}\gamma^{(0)} \ , \quad \gamma^{l}=\frac{w}{a}\left( {\begin{array}{cc}
			\sigma^{l} & 0 \\
			0 & -\sigma^{l}\\
	\end{array} } \right) \ , \quad
	\gamma^{4}=i\frac{w}{a}\left( {\begin{array}{cc}
			0 & 1 \\
			1 & 0\\
	\end{array} } \right) \  ,
	\label{Dirac-matrices-curved}
\end{equation}
where we have introduced the $2 \times 2$ matrices for $l=(r,\phi,w)$
\begin{equation}
	\sigma^{r}=\left( {\begin{array}{cc}
			0 & e^{-iq\phi} \\
			e^{iq\phi} & 0\\
	\end{array} } \right) \ , \ \sigma^{\phi}=-\frac{i}{r}\left( {\begin{array}{cc}
			0 & e^{-iq\phi} \\
			-e^{iq\phi} & 0\\
	\end{array} } \right) \ , \ \sigma^{w}=\left( {\begin{array}{cc}
			1 & 0 \\
			0 & -1\\
	\end{array} } \right) \ .
	\label{Pauli-matrices-curved}
\end{equation}
For the spin connection we obtain \cite{Bellucci:2020byz}
\begin{equation}
	\Gamma_{\mu}=\frac{1}{2a}\gamma^{(3)}\gamma_{\mu}+\frac{(1-q)}{2}\gamma^{(1)}\gamma^{(2)}\delta^{\phi}_{\mu} \ , \quad \Gamma_{z}=0   \  .
\end{equation}
Its contraction with the Dirac matrices reads,
\begin{equation}
	\gamma^{\mu}\Gamma_{\mu}=-\frac{2}{w}\gamma^{w}+\frac{1-q}{2r}\gamma^{r}.
\end{equation}
For this case the Dirac equation takes the following form:
\begin{equation}
	\label{Dirac_eq_1}
	\bigg(\gamma^{\mu}(\partial_{\mu}+ieA_{\mu})	-\frac{2}{w}\gamma^{w}+\frac{1-q}{2r}\gamma^{r}+i\tilde{s}m\bigg)\psi=0  \   .
\end{equation}
Let us also assume the vector potential is the same as given in  \eqref{VP}. 

Assuming the following Ansatz for the positive energy solution of the Dirac equation,
\begin{eqnarray}
\psi=\left( {\begin{array}{c}
			\varphi \\
		\chi \\
	\end{array} } \right)e^{-i(Et-kz)} \ ,
	\label{psi_func}
\end{eqnarray} 
where $\varphi$ and $\chi$ correspond to two-component spinors. The Dirac equation \eqref{Dirac_eq_1} can be written as,
\begin{eqnarray}
	\bigg(\sigma^{l}\partial_{l}-\frac{2}{w}\sigma^{w}+\frac{1-q}{2r}\sigma^{r}+ieA_{\phi}\sigma^{\phi}-\frac{{s}ma}{w}\bigg)\varphi-i(E+k)\chi&=&0, \nonumber \\
	\bigg(\sigma^{l}\partial_{l}-\frac{2}{w}\sigma^{w}+\frac{1-q}{2r}\sigma^{r}+ieA_{\phi}\sigma^{\phi}+\frac{{s}ma}{w}\bigg)\chi+i(k-E)\varphi&=&0 \ .
	\label{dub-diff-eq}
\end{eqnarray}
Substituting the function $\chi$ from the second equation into the first one, we get the following second order differential equation for the $\varphi$:
\begin{eqnarray}
	&&\bigg\{\partial_{r}^{2}+\frac{1}{r}\partial_{r}+\frac{1}{r^2}\bigg[\partial_{\phi}+ieA_\phi-\frac{i(1-q)}{2}\sigma_{w}\bigg]^2+\partial_{w}^{2}-\frac{4}{w}\partial_w+\frac{6-({s}ma)^2-(-1)^l{s}ma}{w^2}\nonumber\\&&+(E^2-{k})\bigg\}\varphi=0 \ .
	\label{diff-eq}
\end{eqnarray}

At this point we will assume that the two-component spinor $\phi$ compatible with the cylindrical symmetry of the problem, can be expressed as,
\begin{equation}
	\varphi=\left( {\begin{array}{c}
			C_{1}R_{1}(r)W_{1}(w)e^{iqn_{1}\phi} \\
			C_{2}R_{2}(r)W_{2}(w)e^{iqn_{2}\phi} \\
	\end{array} } \right) \ ,
	\label{upper-comp-1}
\end{equation}
with $C_{1}$ and $C_{2}$ being two arbitrary constants.

Substituting the above expression into \eqref{diff-eq}, we verify that a solution for $R_{l}(r)$, for $l=1, \ 2$, regular on the $r=0$, can be expressed in terms of the Bessel function of the first kind, $R_{l}(r)=J_{\beta_{l}}(\lambda r)$, having its order given by
\begin{equation}
	\label{index}
	\beta_{1}=|q(n_{1}+\alpha)-(1-q)/2| \ \ , \ \ \beta_{2}=|q(n_{2}+\alpha)+(1-q)/2| \ ,
\end{equation}
with $n_{l}=0,\pm1, \pm2,....$ In \eqref{index} we have introduced the notation for $\alpha$ given in \eqref{const}. As to the function associated with the Poincaré coordinate, $W_l(w)$, the general solution is given in terms of a linear combination of the functions $w^{5/2}J_{\nu_l}(pw)$ and $w^{5/2}Y_{\nu_l}(pw)$, where $Y_{\nu}(x)$ is the Neumann function \cite{Abra}, with the order
\begin{equation}\nu_{l}=|sma+(-1)^l/2|.
	\label{rel-nu}
\end{equation}
Considering $ma\ge1/2$, the Neumann function must be excluded according to the normalizability conditions of the mode functions, consequently we will adopt the solution
\begin{equation}
	W_{l}(w)=w^{5/2}J_{\nu_{l}}(pw) \ .
\end{equation}

The energy associated to the modes is given by
\begin{equation}
	E=\sqrt{\lambda^2+p^2+k^2} \ .
\end{equation}
For sake of simplicity, from now on we will assume the representation of the Clifford algebra corresponding to ${s}=1$.

Taking the explicit expression found for $\varphi$ into the first equation in \eqref{dub-diff-eq}, after some intermediate steps, we obtain
\begin{eqnarray}
	\chi=w^{5/2}\left( {\begin{array}{c}
			B_{1}J_{\beta_{1}}(\lambda r)J_{\nu_{1}}(pw)e^{iqn_{1}\phi} \\
			B_{2}J_{\beta_{2}}(\lambda r)J_{\nu_{2}}(pw)e^{iqn_{2}\phi} \\
	\end{array} } \right)
	\label{lower-spinor}
\end{eqnarray}	
with the relations
\begin{equation}
	n_{2}=n_{1}+1 \ , \ \beta_{2}=\beta_{1}+\epsilon_{n_{1}},
\end{equation}
being $\epsilon_{n}=1$ for $n>-\alpha$ and $\epsilon_{n}=-1$ for $n<-\alpha$. The coefficients $B_{1,2}$ in \eqref{lower-spinor} are given in terms of the coefficients $C_1$ and $C_2$, by
\begin{equation}
	B_{1}=\frac{i}{E+{k}}(pC_{1}-\epsilon_{n}\lambda C_{2}) \ , \ B_{2}=\frac{i}{E+{k}}(\epsilon_{n}\lambda C_{1}+pC_{2}) \ .
	\label{coeff-relations}
\end{equation}

Substituting the upper and lower two-component spinors into \eqref{psi_func}, this wave function is an eigenfunction of the total angular momentum projected along the direction of the string
\begin{equation}
	\hat{J}_{w}\psi=\bigg(-i\partial_{\phi}+\frac{q}{2}\Sigma^w\bigg)\psi \ , \ \Sigma^w=\left( {\begin{array}{cc}
			\sigma^{w} & 0 \\
			0 & \sigma^{w}\\
	\end{array} } \right) \ ,
\end{equation}
with
\begin{equation}
	j=n_{1}+1/2 \ , \ j=\pm 1/2, \pm 3/2,... \ .
\end{equation}

At this point we should mention that the components of the fermionic wave function obtained, due to relation \eqref{coeff-relations}, contains two independent coefficients. By imposing the normalization condition on the wave function an additional relation between these two constants is obtained. Thus, one of the coefficients remains arbitrary. In order to determine this coefficient it is necessary the imposition of some additional condition on the coefficients. The imposition of this condition comes from the fact that the set of quantum numbers $\{\lambda,p,j,k\}$ do not specify the fermionic wave function uniquely. Consequently  an additional quantum number is required.

In order to specify the second constant we will adopt the following relation between the upper and lower components \cite{Bordag}:
\begin{equation}
	\chi_{1}=\kappa\varphi_{1}, \ \chi_{2}=-\varphi_{2}/\kappa.
	\label{c-coeff}
\end{equation}
From the expressions for the spinor components we find the eigenvalues of the parameter $\kappa$,
\begin{equation}
	\kappa=\kappa_{\eta}=\frac{-{k}+\eta\sqrt{{k}^{2}+p^2}}{p}, \ \eta=\pm1,
\end{equation}
and the relation
\begin{equation}
	C_{2}=-\frac{\epsilon_{n}\kappa_{\eta}}{\lambda}\big(E-\eta\sqrt{{k}^2+p^2}\big)C_1,
\end{equation}
for the coefficients $C_l$. Now the fermionic wave function is uniquely specified by the set of quantum numbers $\sigma=(\lambda,p,j,k,\eta)$. 

Now taking into account all the above considerations, the positive-energy fermionic wave function can be expressed as
\begin{equation}
	\psi^{(+)}_{\sigma}(x)=C^{(+)}_{\sigma}e^{-i(Et-kz)}w^{5/2}\left( {\begin{array}{c}
			J_{\beta_{j}}(\lambda r)J_{\nu_{1}}(pw) \\
			-\epsilon_{j}\kappa_{\eta}b^{(+)}_{\eta}J_{\beta_{j}+\epsilon_{j}}(\lambda r)J_{\nu_{2}}(pw)e^{iq\phi} \\
			i\kappa_{\eta}J_{\beta_{j}}(\lambda r)J_{\nu_{2}}(pw)\\
			i\epsilon_{j}b^{(+)}_{\eta}J_{\beta_{j}+\epsilon_{j}}(\lambda r)J_{\nu_{1}}(pw)e^{iq\phi}
	\end{array} } \right)e^{iq(j-1/2)\phi} \ ,
	\label{positive-energy-wfunc}
\end{equation}
where $\epsilon_{j}=1$ for $j>-\alpha$ and $\epsilon_{j}=-1$ for $j<-\alpha$, and the order of the Bessel function is defined as
\begin{equation}
	\label{beta}
	\beta_{j}=q|j+\alpha|-\epsilon_{j}/2.
\end{equation}

In \eqref{positive-energy-wfunc} we introduced the notation
\begin{equation}
	b^{(\pm)}_{\eta}=\frac{E\mp \eta\sqrt{{k}^2+p^2}}{\lambda}.
\end{equation}
Note that one has the relation $b^{(+)}_{\eta}b^{(-)}_{\eta}=1$.

Finally the coefficient $C^{(+)}_{\sigma}$ is determined by the normalization condition
\begin{equation}
	\int d^3x\sqrt{\gamma}(\psi^{(+)}_{\sigma})^\dagger\psi^{(+)}_{\sigma^\prime}=\delta_{\sigma,\sigma^\prime} \ , \
	\label{norm-cond}
\end{equation}
where $\gamma$ is the determinant of the spatial metric. The delta symbol on the right-hand side is understood
as the Dirac delta function for continuous quantum numbers $(\lambda, \  p, \ k)$ and the Kronecker delta for discrete ones $(j, \ \eta)$. Taking the eigenspinor in \eqref{positive-energy-wfunc} into \eqref{norm-cond} and using the value of the standard integral involving the products of the Bessel functions \cite{Grad}, we find
\begin{equation}
	\label{C+}
	|C^{(+)}_{\sigma}|^{2}=\frac{\eta qp^2\lambda^2}{16\pi^2a^4E\kappa_{\eta}b^{(+)}_{\eta}
		\sqrt{{k}^2+p^2}} \ .
\end{equation}

For the negative-energy fermionic mode function a similar procedure can be adopted, providing the expression
\begin{equation}
	\psi^{(-)}_{\sigma}(x)=C^{(-)}_{\sigma}e^{i(Et+ikz)}w^{5/2}\left( {\begin{array}{c}
			J_{\beta_{j}}(\lambda r)J_{\nu_{1}}(pw) \\
			\epsilon_{j}\kappa_\eta b^{(-)}_{\eta}J_{\beta_{j}+\epsilon_{j}}(\lambda r)J_{\nu_{2}}(pw)e^{iq\phi} \\
			i\kappa_{\eta}J_{\beta_{j}}(\lambda r)J_{\nu_{2}}(pw)\\
			-i\epsilon_{j}b^{(-)}_{\eta}J_{\beta_{j}+\epsilon_{j}}(\lambda r)J_{\nu_{1}}(pw)e^{iq\phi}
	\end{array} } \right)e^{iq(j-1/2)\phi} \ ,
	\label{negative-energy-wfunc}
\end{equation}
and the normalization constant is given by the relation
\begin{equation}
	\label{C-}
	|C^{(-)}_{\sigma}|^{2}=\frac{\eta qp^2\lambda^2}{16\pi^2 a^4E\kappa_{\eta}b^{(-)}_{\eta}\sqrt{{k}^2+p^2}} \ .
\end{equation}

\subsection{Fermionic condensate}
\label{condensate}

The fermion condensate (FC) is defined as the vacuum expectation value $\langle 0|\bar{\psi}\psi |0\rangle \equiv\langle \bar{\psi}\psi \rangle $, being $|0\rangle $ the vacuum state and $\bar{\psi}=\psi ^{\dagger }\gamma ^{(0)}$ is the Dirac adjoint.\footnote{Note that in the definition of the Dirac adjoint $\gamma ^{(0)}$ is the flat spacetime matrix  \eqref{gamma_matrices_flat}}.  In order to calculate the FC
we expand the field operator in terms of the complete set of the positive and negative energy fermionic modes $\{\psi _{\sigma }^{(+)},\psi _{\sigma }^{(-)}\} $. Using the anticommutation relations for the creation and annihilation operators, we get \cite{Bellucci:2021zyf}:
\begin{equation}
	\langle \bar{\psi}\psi \rangle =-\frac{1}{2}\sum_{\sigma }\sum_{\chi
		=-,+}\chi \bar{\psi}_{\sigma }^{(\chi )}\psi _{\sigma }^{(\chi )}  \  .
	\label{FC}
\end{equation}
In \eqref{FC} the summation goes over the complete set of quantum numbers as
\begin{equation}
	\sum_{\sigma }=\sum_{j}\int_{0}^{\infty }d\lambda \int_{0}^{\infty
	}dp\int_{-\infty }^{\infty }dk\sum_{\eta =\pm 1}\ ,  \label{Sumsig}
\end{equation}
with $\sum_{j}=\sum_{j=\pm 1/2,\pm 3/2,\cdots }$.

 The operators in the definition of the FC are evaluated at the same spacetime point
and, consequently,  the expression in the right-hand side of (\ref{FC}) is divergent.
In order to obtain a finite and well defined quantity we have to regularize it. Several regularization procedures can be employed. As we will see the details of the procedure adopted to calculate the FC is not relevant in what follows.

Substituting the fermionic mode functions in (\ref{FC}), for the FC we obtain
\begin{equation}
	\langle \bar{\psi}\psi \rangle _{\mathrm{cs}}^{\mathrm{AdS}}=-\frac{qw^{5}}{%
		16\pi ^{2}a^{4}}\sum_{\sigma }\sum_{\chi =-,+}\frac{\chi \eta p^{2}\lambda
		^{2}}{E\sqrt{p^{2}+k_{z}^{2}}}J_{\nu _{1}}(pw)J_{\nu _{2}}(pw)\left[ b_{\eta
	}^{(-\chi )}J_{\beta _{j}}^{2}(\lambda r)-b_{\eta }^{(\chi )}J_{\beta
		_{j}+\epsilon _{j}}^{2}(\lambda r)\right] .  \label{FC2}
\end{equation}
We have used in \eqref{FC2} the property $b_{\eta }^{(+)}b_{\eta }^{(-)}=1$. Taking the expression for $b_{\eta }^{(\pm )}$ and summing over $\eta $, this formula can be simplified to
\begin{eqnarray}
	\langle \bar{\psi}\psi \rangle _{\mathrm{cs}}^{\mathrm{AdS}} &=&-\frac{qw^{5}}{4\pi ^{2}a^{4}}\sum_{j}\int_{0}^{\infty }d\lambda \int_{0}^{\infty
	}dp\int_{-\infty }^{\infty }dk_{z}\frac{\lambda p^{2}}{\sqrt{\lambda
			^{2}+p^{2}+k_{z}^{2}}}  \notag \\
	&&\times J_{\nu _{1}}(pw)J_{\nu _{2}}(pw)\left[ J_{\beta _{j}}^{2}(\lambda
	r)+J_{\beta _{j}+\epsilon _{j}}^{2}(\lambda r)\right] \ .  \label{AdScs}
\end{eqnarray}
In fact the FC presents opposite signs for the fields with $s=+1 $ and $s=-1$. However, as we have already mentioned we consider the case $s=+1$. 

To obtain a more workable expression for (\ref{AdScs}), we use the identity
\begin{equation}
	\frac{1}{\sqrt{\lambda ^{2}+p^{2}+k_{z}^{2}}}=\frac{2}{\sqrt{\pi }}%
	\int_{0}^{\infty }d\tau e^{-\tau ^{2}(\lambda ^{2}+p^{2}+k_{z}^{2})}\ .
	\label{first-id}
\end{equation}

The substitution of the above identity into (\ref{AdScs}) allows us to integrate over the quantum numbers $\lambda $, $p$ and $k$ by using the results from \cite{Grad}. After some intermediate steps, and introducing a new integration
variable $y$ defined by $y=r^{2}/2\tau ^{2}$, we get
\begin{eqnarray}
	\langle \bar{\psi}\psi \rangle _{\mathrm{cs}}^{\mathrm{AdS}} &=&-\frac{%
		q(w/r)^{6}}{4\pi ^{2}a^{4}}\int_{0}^{\infty }dx\,x^{2}e^{-(1+\rho ^{-2})x}%
	\left[ I_{\nu _{1}}\left( x/\rho ^{2}\right) -I_{\nu _{2}}\left( x/\rho
	^{2}\right) \right]  \notag \\
	&&\times \sum_{j}\left[ I_{\beta _{j}}\left( x\right) +I_{\beta
		_{j}+\epsilon _{j}}\left( x\right) \right] \ ,  \label{FCcs1}
\end{eqnarray}%
where $I_{\nu }(z)$ represents the modified Bessel function \cite{Abra}. In (\ref{FCcs1}),
\begin{equation}
	\rho =r/w  \label{rho}
\end{equation}%
is the proper distance from the string, $r_{p}=ar/w$, measured in units of $%
a $.

The parameter $\alpha $ presents in the definition of $\beta_j$ in \eqref{beta} can be written as $\alpha =\alpha _{0}+n_{0}$, being $n_0$ an integer number and $|\alpha _{0}|<1/2$. Redefining  $j+n_{0}\rightarrow j$, we see that the VEVs do not depend on $n_{0}$. This is a typical Aharanov-Bohm effect present in this analysis. From now on we will take in our discussion $\alpha=\alpha _{0}$ without loss of generality.

 A more workable expression for
\begin{equation}
	{\mathcal{J}}(q,\alpha _{0},x)=\sum_{j}\left[ I_{\beta _{j}}\left( x\right)
	+I_{\beta _{j}+\epsilon _{j}}\left( x\right) \right]  \label{Jcal}  \  ,
\end{equation}
was obtained in \cite{Beze10f3}. It reads,
\begin{eqnarray}
	{\mathcal{J}}(q,\alpha _{0},x) &=&\frac{2}{q}e^{x}+\frac{4}{q}%
	\sideset{}{'}{\sum}_{k=1}^{[q/2]}(-1)^{k}c_{k}\cos (2\pi k\alpha
	_{0})e^{x\cos (2\pi k/q)}  \notag \\
	&&+\frac{4}{\pi }\int_{0}^{\infty }du\frac{h(q,\alpha _{0},u)e^{-x\cosh 2u}}{%
		\cosh (2qu)-\cos (q\pi )},  \label{Jf}
\end{eqnarray}%
where $[q/2]$ represents the integer part of $q/2$ and the prime on the summation
sign means that for even values of $q$ the term with $k=q/2$ should be
halved. In the case $1\leqslant q<2$, the second term on the right-hand side of (\ref{Jf}) must be discarded. Note that $h(q,\alpha _{0},z)$ and ${\mathcal{J}}(q,\alpha _{0},y)$ are even functions of $\alpha _{0}$.

Moreover, in (\ref{Jf}), we have introduced the notations $c_{k}=\cos (\pi k/q)$ and 
\begin{equation}
	h(q,\alpha _{0},u)=\sinh u\sum_{\chi =+,-}\cos [\pi q(1/2+\chi \alpha
	_{0})]\sinh [(1-2\chi \alpha _{0})qu]  \label{hq} \  .
\end{equation}
In the special case $\alpha _{0}=0$ (the magnetic flux along the string is a multiple of the flux quantum) the expression (\ref{hq}) is simplified to
\begin{equation}
	h(q,0,u)=2\cos (\pi q/2)\sinh (qu)\sinh u,  \label{Hq0}
\end{equation}%
and in (\ref{Jf}) the integral term vanishes for odd values of $q$.

As in the case of the VEV of the scalar field squared, analyzed in subsection \ref{Field_squared}, the FC also can be decomposed as,
\begin{equation}
\langle \bar{\psi}\psi \rangle
_{\mathrm{cs}}^{\mathrm{AdS}}=\langle \bar{\psi}\psi \rangle ^{\mathrm{AdS}}+	\langle \bar{\psi}\psi \rangle _{\mathrm{cs}} \  .
	\label{FC_CS_Ren}
\end{equation}

Substituting \eqref{Jf} into \eqref{FCcs1}, the contribution coming from the term with provides 
\begin{equation}
	\langle \bar{\psi}\psi \rangle ^{\mathrm{AdS}}=-\frac{1}{2\pi ^{2}a^{4}}%
	\int_{0}^{\infty }dx\,x^{2}e^{-x}\left[ I_{\nu _{1}}\left( x\right) -I_{\nu
		_{2}}\left( x\right) \right] \  .  \label{FCAdS}
\end{equation}%
This contribution does not depend on $q$ and $\alpha _{0}$ and corresponds to th FC in a pure
$(1+4)-$dimensional AdS spacetime. Due to the maximal symmetry of the AdS spacetime the latter does not depend on spacetime point. The expression in the right-hand side of (\ref{FCAdS}) is divergent and needs a regularization with further
renormalization. Because we are interested in the effects
induced by cosmic string, the corresponding contribution to the FC is:
\begin{eqnarray}
	\langle \bar{\psi}\psi \rangle _{\mathrm{cs}} &=&-\frac{1}{\pi ^{2}a^{4}}%
	\int_{0}^{\infty }dx\,x^{2}e^{-x}\left[ I_{\nu _{1}}\left( x\right) -I_{\nu
		_{2}}\left( x\right) \right]  \notag \\
	&&\times \Biggl\{\sideset{}{'}{\sum}_{k=1}^{[q/2]}(-1)^{k}\cos (\pi k/q)\cos
	(2\pi k\alpha _{0})e^{-2x\rho ^{2}\sin ^{2}(\pi k/q)}  \notag \\
	&&+\frac{q}{\pi }\int_{0}^{\infty }du\frac{h(q,\alpha _{0},2u)\sinh u}{\cosh
		(2qu)-\cos (q\pi )}e^{-2x\rho ^{2}\cosh ^{2}u}\Biggr\}\ .  \label{FCcs0}
\end{eqnarray}%
The integral over $x$ can be evaluated by using the integration formula from \cite{Grad}:
\begin{equation}
	\int_{0}^{\infty }dx\,x^{\mu -1}e^{-ux}I_{\nu }(x)=\sqrt{\frac{2}{\pi }}%
	\frac{e^{-(\mu -1/2)\pi i}Q_{\nu -1/2}^{\mu -1/2}(u)}{(u^{2}-1)^{\mu /2-1/4}} 
	\ ,  \label{function_Q}
\end{equation}%
with $u>1$ and $Q_{\nu }^{\mu }(z)$ represents the associated Legendre
function \cite{Abra}. After some intermediate steps the Eq. \eqref{FCcs0} can be expressed by
\begin{eqnarray}
	\langle \bar{\psi}\psi \rangle _{\mathrm{cs}} &=&-\frac{\sqrt{2}}{\pi
		^{5/2}a^{4}}\Biggl[\sideset{}{'}{\sum}_{k=1}^{[q/2]}(-1)^{k}\cos (\pi
	k/q)\cos (2\pi k\alpha _{0}){\mathcal{Z}}_{ma}(u_{k})  \notag \\
	&&+\frac{q}{\pi }\int_{0}^{\infty }dx\frac{h(q,\alpha _{0},2x)\sinh x}{\cosh
		(2qx)-\cos (q\pi )}{\mathcal{Z}}_{ma}(u_{x})\Biggr]\ .  \label{FC-cs}
\end{eqnarray}%
Where we have introduced the notation
\begin{equation}
	{\mathcal{Z}}_{ma}(u)=F_{\nu _{1}}(u)-F_{\nu _{2}}(u)\,,  \label{Z_function}
\end{equation}
being the function%
\begin{equation}
	F_{\nu }(u)=\frac{e^{-i5\pi /2}Q_{\nu -1/2}^{5/2}(u)}{(u^{2}-1)^{5/4}}\,,
	\label{Fnu}
\end{equation}%
and the variables
\begin{eqnarray}
	u_{k} &=&1+2\rho ^{2}\sin ^{2}(\pi k/q)\ ,  \notag \\
	u_{x} &=&1+2\rho ^{2}\cosh ^{2}x\ .  \label{args-cs}
\end{eqnarray}%
Note that the FC $\langle \bar{\psi}\psi \rangle _{\mathrm{cs}}$ depends on
the coordinates $r$ and $w$ through the ratio (\ref{rho}). This property is
a consequence of the maximal symmetry of the AdS spacetime.

For the case of massless field, the function ${\mathcal{Z}}_{ma}(u)$ is expressed in
terms of elementary functions:
\begin{equation}
	{\mathcal{Z}}_{0}(u)=\frac{3\sqrt{\pi }}{4(1+u)^{5/2}}\ .  \label{massless-Z}
\end{equation}%
Taking this expression into (\ref{FC-cs}), we get
\begin{eqnarray}
	\langle \bar{\psi}\psi \rangle _{\mathrm{cs}} &=&-\frac{3}{16\pi ^{2}a^{4}}%
	\Biggl[\sideset{}{'}{\sum}_{k=1}^{[q/2]}(-1)^{k}\frac{\cos (\pi k/q)\cos
		(2\pi k\alpha _{0})}{(1+\rho ^{2}\sin ^{2}(\pi k/q))^{5/2}}  \notag \\
	&&+\frac{q}{\pi }\int_{0}^{\infty }dx\frac{\sinh x}{\cosh (2qx)-\cos (q\pi )}%
	\frac{h(q,\alpha _{0},2x)}{(1+\rho ^{2}\cosh ^{2}x)^{5/2}}\Biggr]\ .
	\label{FCm0}
\end{eqnarray}

\begin{figure}[tbph]
	\begin{center}
		\begin{tabular}{cc}
			\epsfig{figure=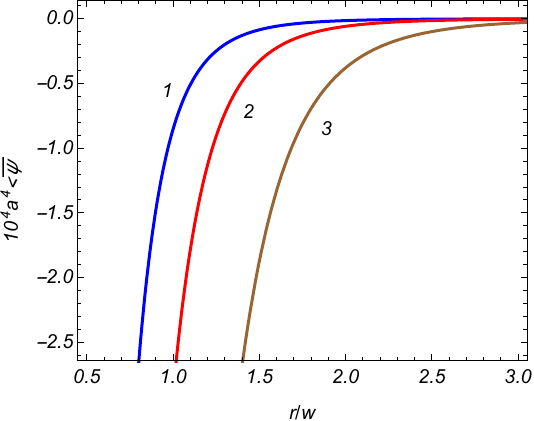,width=7.cm,height=5.5cm} & \quad %
			\epsfig{figure=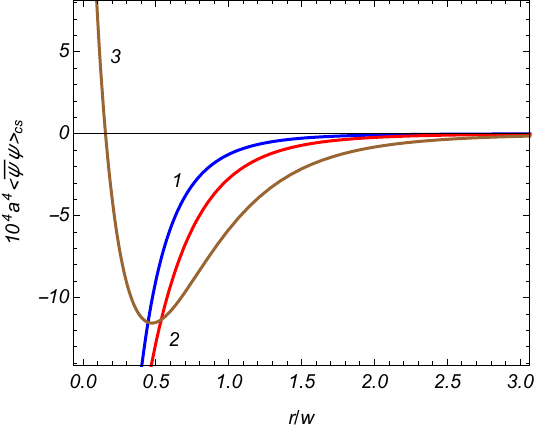,width=7.cm,height=5.5cm}%
		\end{tabular}%
	\end{center}
	\caption{The FC $\langle \bar{\protect\psi}\protect\psi \rangle _{\mathrm{cs}%
		}$ as a function of $r/w$ for massive field (left graph), considering $ma=1$,
		and massless one (right graph). The graphs are plotted for $\protect%
		\alpha _{0}=0.3$ and for different values of $q$ (numbers near the curves).}
	\label{fig1a}
\end{figure}

In Fig. \ref{fig1a} we exhibit the behavior of FC as function of $r/w$. The graphs are plotted  taking $\alpha _{0}=0.3$. The numbers near the curves correspond the values of the
parameter $q$. The left plot is for a massive fermionic field with $ma=1$ and the right one is for massless field.

\subsection{The VEV of the Energy-Momentum tensor}
\label{Fermionic_energy-momentum}
In this section we investigate topological effects of a cosmic string on the vacuum expectation value (VEV) of the energy-momentum tensor associated to a charged  fermionic quantum field in $(4+1)-$dimensional locally AdS spacetime.

The VEV of the energy-momentum tensor associated with fermionic field can be evaluated by using the mode-sum formula below \cite{Bellucci:2021jch},
\begin{equation}
	\left\langle T_{\mu \nu }\right\rangle =-\frac{i}{4}\sum_{\sigma }\sum_{\chi
		=-,+}\chi {\left[ \bar{\psi}_{\sigma }^{(\chi )}\gamma _{(\mu }\mathcal{D}%
		_{\nu )}\psi _{\sigma }^{(\chi )}-(\mathcal{D}_{(\mu }\bar{\psi}_{\sigma
		}^{(\chi )})\gamma _{\nu )}\psi _{\sigma }^{(\chi )}\right] }\ ,  \label{EMT}
\end{equation}
where $\{\psi _{\sigma }^{(+)},\psi _{\sigma }^{(-)}\}$ represents the complete set
of the positive and negative energy fermionic modes.  In the expression above, the brackets in the index expression mean the symmetrization over the enclosed indices. Moreover, the Dirac adjoint is defined as usual $\bar{\psi}_{\sigma }^{(\chi )}=\psi _{\sigma }^{(\chi
	)\dagger }\gamma ^{(0)}$ and $\mathcal{D}_{\mu }{\bar{\psi}}_{\sigma
}^{(\chi )}=\partial _{\mu }{\bar{\psi}}_{\sigma }^{(\chi )}-ieA_{\mu }{\bar{%
		\psi}}_{\sigma }^{(\chi )}-{\bar{\psi}}_{\sigma }^{(\chi )}\Gamma _{\mu }$
for $\chi =+,-$. 

Having the normalized fermionic mode functions, \eqref{positive-energy-wfunc} and \eqref{negative-energy-wfunc}, with their respective coefficients \eqref{C+} and \eqref{C-}, the VEV of the energy-momentum tensor is evaluated by using the formula (\ref{EMT}), where
the summation over $\sigma $ was defined in \eqref{Sumsig}.

After a long calculation the VEVs of the diagonal components are presented in a compact formula as (no summation over $\mu $),
\begin{equation}
	\left\langle T_{\mu }^{\mu }\right\rangle _{\mathrm{cs}}^{\mathrm{AdS}}=-%
	\frac{qw^{6}}{4\pi ^{2}a^{5}}\sum_{j}\int_{0}^{\infty }d\lambda \lambda
	\int_{0}^{\infty }dp\,p\int_{0}^{\infty }dk\,E^{2\delta _{0\mu
		}-1}\left( k_{z}^{2}\right) ^{\delta _{4\mu }}R_{\beta _{j}}^{(\mu )}\left(
	\lambda r\right) W_{\nu }^{(\mu )}(pw)  \  ,  \label{Tkk}
\end{equation}
with $\nu =ma-1/2$. In \eqref{Tkk} we used the notation,
\begin{eqnarray}
	R_{\beta _{j}}^{(0)}\left( \lambda r\right) &=&R_{\beta _{j}}^{(3)}\left(
	\lambda r\right) =-R_{\beta _{j}}^{(4)}\left( \lambda r\right) =J_{\beta
		_{j}}^{2}(\lambda r)+J_{\beta _{j}+\epsilon _{j}}^{2}(\lambda r),  \notag \\
	R_{\beta _{j}}^{(1)}\left( \lambda r\right) &=&\epsilon _{j}\lambda ^{2} 
	\left[ J_{\beta _{j}}^{\prime }(\lambda r)J_{\beta _{j}+\epsilon
		_{j}}(\lambda r)-J_{\beta _{j}}(\lambda r)J_{\beta _{j}+\epsilon
		_{j}}^{\prime }(\lambda r)\right] ,  \notag \\
	R_{\beta _{j}}^{(2)}\left( \lambda r\right) &=&-\frac{\lambda }{r}(2\beta
	_{j}+\epsilon _{j})J_{\beta _{j}}(\lambda r)J_{\beta _{j}+\epsilon
		_{j}}(\lambda r),  \label{R2}
\end{eqnarray}%
and%
\begin{eqnarray}
	W_{\nu }^{(\mu )}(pw) &=&J_{\nu }^{2}(pw)+J_{\nu +1}^{2}(pw),\;\mu =0,1,2,4,
	\notag \\
	W_{\nu }^{(3)}(pw) &=&p^{2}\left[ J_{\nu }^{\prime }(pw)J_{\nu
		+1}(pw)-J_{\nu }(pw)J_{\nu +1}^{\prime }(pw)\right]  \ .  \label{W3}
\end{eqnarray}
The off-diagonal components of the vacuum energy-momentum tensor vanish.

As to the scalar case, the expression given in \eqref{Tkk} can be decomposed in the form,
\begin{eqnarray}
	\label{decomposition}
		\left\langle T_{\mu }^{\mu }\right\rangle _{\mathrm{cs}}^{\mathrm{AdS}}=\left\langle T_{\mu \nu }\right\rangle^{AdS}+\left\langle T_{\mu \nu }\right\rangle_{cs} \ , 
\end{eqnarray}
where the first term on the right-hand side represents corresponds to the pure AdS part, and the second is the correction induced by the presence of a cosmic string. Here in this paper we are interested in calculating $\left\langle T_{\mu \nu }\right\rangle_{cs}$. In fact the $\left\langle T_{\mu \nu }\right\rangle_{AdS}$ has already been calculated in \cite{Camporesi:1992wn} by using the zeta-function technique.

In order to extract the contribution induced by the cosmic string in (\ref{Tkk}), we use the integral representation below for the term associated with the energy, $E$:
\begin{equation}
	E^{2\delta _{0\mu }-1}=\frac{2}{\sqrt{\pi }}\int_{0}^{\infty }d\tau \,\left(
	\partial _{-\tau ^{2}}\right) ^{\delta _{0\mu }}e^{-(\lambda
		^{2}+p^{2}+k_{z}^{2})\tau ^{2}}.  \label{Id}
\end{equation}%
In this way the integral over $k$ can be evaluated trivially. As to the integral over $p$ we use the the formula \cite{Grad}
\begin{equation}
	\int_{0}^{\infty }dppe^{-p^{2}\tau ^{2}}W_{\nu }^{(\mu )}(pw)=\frac{\left(
		-1\right) ^{\delta _{3\mu }}e^{-x}}{(2\tau ^{2})^{1+\delta _{3\mu }}}\left[
	I_{\nu }(x)+I_{\nu +1}(x)\right] _{x=w^{2}/2\tau ^{2}},  \label{Intp}
\end{equation}%
where $I_{\nu }(x)$ represents the modified Bessel function \cite{Abra}. Finally the
integral over $\lambda $ involving the functions $R_{\beta _{j}}^{(\mu
	)}\left( \lambda r\right) $ with $\mu \neq 2$ has similar structure as 
\eqref{Intp} and the integral for $R_{\beta _{j}}^{(2)}\left( \lambda
r\right) $ is evaluated by using the relation
\begin{equation}
	R_{\beta _{j}}^{(1)}\left( \lambda r\right) =\lambda ^{2}R_{\beta
		_{j}}^{(0)}\left( \lambda r\right) +R_{\beta _{j}}^{(2)}\left( \lambda
	r\right) .  \label{R1}
\end{equation}
In this way, the components with $\mu =0,1,3,4$ are presented as (no summation over $\mu $)
\begin{equation}
	\left\langle T_{\mu }^{\mu }\right\rangle _{\mathrm{cs}}^{\mathrm{AdS}}=
	\frac{qa^{-5}}{8\pi ^{2}}\int_{0}^{\infty }dx\,x^{2}e^{-x(1+\rho ^{2})}\left[
	I_{\nu }(x)+I_{\nu +1}(x)\right] {\mathcal{J}}(q,\alpha _{0},x\rho ^{2}) \  ,
	\label{Tkk2}
\end{equation}%
with ${\mathcal{J}}(q,\alpha ,y)$, being given in \eqref{Jf}

It can be shown that  $	\left\langle T_{2 }^{2 }\right\rangle _{\mathrm{cs}}^{\mathrm{AdS}}$  can be obtained by using the relation below,
\begin{equation}
	\langle T_{2}^{2}\rangle _{\mathrm{cs}}^{\mathrm{AdS}}=\left( 1+r\partial
	_{r}\right) \langle T_{0}^{0}\rangle _{\mathrm{cs}}^{\mathrm{AdS}}\ .
	\label{T22}
\end{equation}

Substituting (\ref{Jf}) into (\ref{Tkk2}), the VEV given by \eqref{Tkk}, can be expressed as given in \eqref{decomposition}.

The first contribution, $\left\langle T_{\mu }^{\mu }\right\rangle ^{\mathrm{AdS}}$, comes from the first term of \eqref{Jf}. It reads,
\begin{equation}
	\left\langle T_{\mu }^{\mu }\right\rangle ^{\mathrm{AdS}}=\frac1{4\pi^2 a^5}\int_{0}^{\infty }dx\,x^{2}e^{-x}\left[ I_{\nu }(x)+I_{\nu +1}(x)\right]
	,  \label{TkkAdS}
\end{equation}%
for all values of $\mu $. We can see that it is independent of a spacetime position, in fact it represents the contribution of the AdS spacetime in the absence of cosmic string, and, as we have already mentioned, it was analyzed in \cite{Camporesi:1992wn}.

The contribution $\left\langle T_{\mu }^{\mu}\right\rangle _{\mathrm{cs}}$ comes from the second and third terms in the right-hand side of (\ref{Jf}) and for the components with 
$\mu =0,1,3,4$ it is presented as 
\begin{equation}
	\left\langle T_{\mu }^{\mu }\right\rangle _{\mathrm{cs}}=\frac{a^{-5}}{\sqrt{%
			2}\pi ^{5/2}}\left[ \sideset{}{'}{\sum}_{k=1}^{[q/2]}(-1)^{k}c_{k}\cos (2\pi
	k\alpha _{0}){\mathcal{F}}_{ma}(u_{k})+\frac{q}{\pi }\int_{0}^{\infty }dx%
	\frac{H(q,\alpha _{0},x){\mathcal{F}}_{ma}(u_{x})}{\cosh (2qx)-\cos (q\pi )}%
	\right] \ ,  \label{Tllcs}
\end{equation}%
with the notations 
\begin{eqnarray}
	u_{k} &=&1+2\rho ^{2}s_{k}^{2},\;s_{k}=\sin (\pi k/q)\ ,  \notag \\
	u_{x} &=&1+2\rho ^{2}\cosh ^{2}x\ ,  \label{args-cs}
\end{eqnarray}%
and%
\begin{equation}
	{\mathcal{F}}_{ma}(u)=\sqrt{\frac{\pi }{2}}\int_{0}^{\infty }dx\,x^{2}e^{-ux}%
	\left[ I_{\nu }(x)+I_{\nu +1}(x)\right] \,.  \label{Ffunc}
\end{equation}%
The component $\left\langle T_{2}^{2}\right\rangle _{\mathrm{cs}}$ is
obtained from 
\begin{equation}
	\langle T_{2}^{2}\rangle _{\mathrm{cs}}=\left( 1+r\partial _{r}\right)
	\langle T_{0}^{0}\rangle _{\mathrm{cs}}.  \label{T22cs}
\end{equation}%
The observable $\left\langle T_{\mu }^{\mu }\right\rangle _{\mathrm{cs}}$ depends
on the coordinates $r$ and $w$ through the ratio $r/w$. This property is a
consequence of the maximal symmetry of the AdS spacetime. We also can note that $\left\langle T_{\mu }^{\mu }\right\rangle _{\mathrm{cs}}$ is an even
function of the parameter $\alpha _{0}$. In terms of the magnetic flux along
the string core, $\left\langle T_{\mu }^{\mu }\right\rangle _{\mathrm{cs}}$
is an even periodic function of the magnetic flux with the period of the
flux quantum. Moreover $\left\langle T_{\mu }^{\mu }\right\rangle _{\mathrm{cs}}$ is finite for $r>0$.

Another comment about our result it that the VEV of the energy-momentum tensor is diagonal. This property is not a consequence of the problem symmetry. As we have seen in subsection \ref{Energy-momentum}, for a scalar field there $\langle T_{3}^{1}\rangle_{\rm{cs}}\neq0$.

 It can be checked that $\left\langle T_{\mu \nu }\right\rangle _{\mathrm{cs}}$ obeys the conservation condition, $\nabla _{\mu }\left\langle T_{\nu}^{\mu }\right\rangle _{\mathrm{cs}}=0$, and its trace obeys relation $\left\langle T_{\mu }^{\mu }\right\rangle _{\mathrm{cs}}=sm\left\langle \bar{\psi}\psi \right\rangle _{\mathrm{cs}}$. In particular, for a massless field the tensor $\left\langle T_{\mu \nu }\right\rangle _{\mathrm{cs}}$ is traceless.
 
 The conservation codition is reduced to the relations
 \begin{equation}
 	\left\langle T_{2}^{2}\right\rangle _{\mathrm{cs}}=\partial
 	_{r}(r\left\langle T_{1}^{1}\right\rangle _{\mathrm{cs}}),\;\left\langle
 	T_{2}^{2}\right\rangle _{\mathrm{cs}}=\left( 1-w\partial _{w}\right)
 	\left\langle T_{3}^{3}\right\rangle _{\mathrm{cs}}.  \label{ContEq}
 \end{equation}
 These relations directly follow from (\ref{T22cs}) by taking into account
 that $\left\langle T_{1}^{1}\right\rangle _{\mathrm{cs}}=\left\langle
 T_{3}^{3}\right\rangle _{\mathrm{cs}}=\left\langle T_{0}^{0}\right\rangle _{ 	\mathrm{cs}}$ and the property that the VEV depends on $r$ and $w$ through the ratio $r/w$.
 
 In the absence of cosmic string, i.e., for the case where $q=1$ and $\alpha _{0}=0$ the corresponding geometry reduces to pure AdS spacetime. Due to the maximal symmetry of this space one has $\left\langle T_{\mu \nu }\right\rangle ^{\mathrm{AdS}}=\mathrm{const}\cdot g_{\mu \nu }$. For $r>0$ the contribution $ \left\langle T_{\mu }^{\mu }\right\rangle _{\mathrm{cs}}$ is finite and the  renormalization is required only for  AdS part, $\left\langle T_{\mu }^{\mu }\right\rangle ^{\mathrm{AdS}}$. This feature is a consequence  of the fact that in the region $r>0$ the local geometrical characteristics  are the same as those in AdS spacetime and, consequently, the divergences are the  same as well. The AdS part $\left\langle T_{\mu \nu }\right\rangle ^{\mathrm{AdS}}$ is widely considered in the literature and here we are most interested in the topological effects induced by the cosmic string and magnetic flux.
 
\begin{figure}[tbph]
 	\begin{center}
 		\epsfig{figure=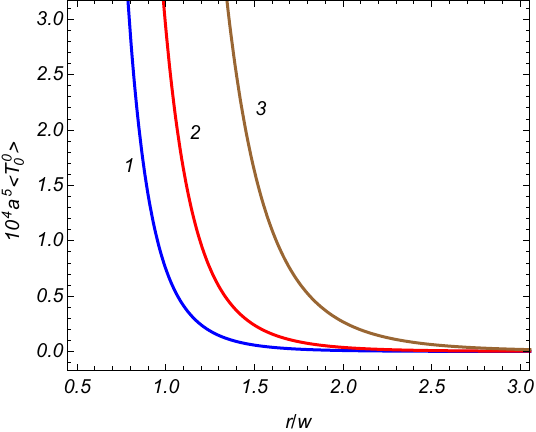,width=7.cm,height=5.5cm}
 	\end{center}
 	\caption{This plot exhibits the behavior of the energy density induced by the cosmic string as function of $r/w$ taken $ma=1$, $\protect\alpha _{0}=0.3$. The 	numbers near the curves correspond the values of $q$. }
 	\label{fig1_a}
 \end{figure}

 To finish this discussion, we present in figure \ref{fig1_a} the plot that exhibits the behavior of the energy density, $\left\langle T_{0}^{0}\right\rangle _{\mathrm{cs}}$  in units of $1/a^{5}$, as function of $r/w$, the proper distance from the string in units of the curvature scale $a$. In our numerical evaluations we have considered $ma=1$ and $\alpha _{0}=0.3$. As to the values of the parameter $q$, they are written near the curves.

\begin{figure}[tbph]
	\begin{center}
		\begin{tabular}{cc}
			\epsfig{figure=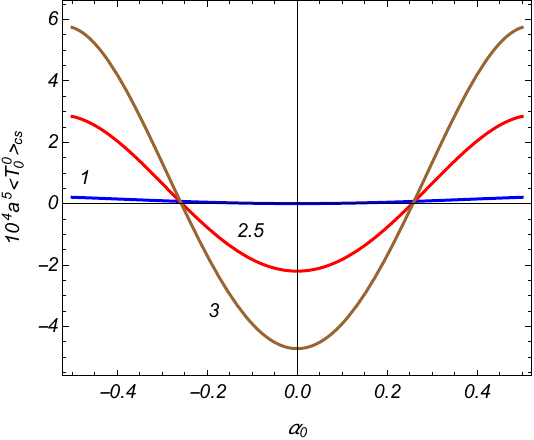,width=7.cm,height=5.5cm} & \quad %
			\epsfig{figure=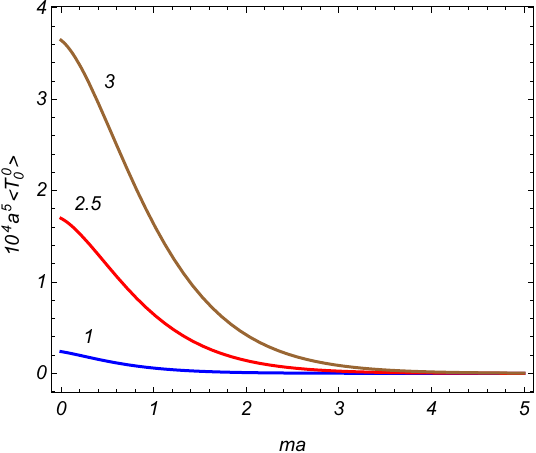,width=7.cm,height=5.5cm}%
		\end{tabular}%
	\end{center}
	\caption{The energy density $\left\langle T_{0}^{0}\right\rangle _{\mathrm{cs%
		}}$ as a function of the parameter $\protect\alpha _{0}$ (left plot) and of
		the mass (right plot) for $r/w=1.5$. On the left plot $ma=0.5$ and on the
		right one $\protect\alpha _{0}=0.3$. The numbers near the curves correspond the
		values assumed for the parameter $q$.}
	\label{fig2_a}
\end{figure}

The figure \ref{fig2_a} presents the dependence of the energy density induced by the string on the parameter $\alpha _{0}$ and on the mass (in units of $1/a$) considering $r/w=1.5$. The left plot is for $ma=0.5$ and for the right plot $\alpha _{0}=0.3$. For both plots the numbers near the curves correspond the values assumed for $q$. As seen, for fixed values of the other parameters, the absolute value of the energy density increases with
increasing planar angle deficit.

\section{Conclude Remarks}
\label{sec4}
In this paper we revisited the analysis of the vacuum polarization associated with a charged bosonic and fermionic fields in a $(D+1)$-dimensional AdS space, with $D\geq 4$.in the presence of a cosmic string carrying  magnetic flux running along its core. Two separate analysis have been done. 

 In section\ref{sec2} we investigate the quantum vacuum associated with bosonic field. Specifically the subsection \ref{Wightman_fun}, we obtain the  positive frequency Wightman function, that is the two-point function used to develop our analysis. In order to do that, by using {\it Popicar\'e} coordinate, we obtain  first the complete set of bosonic modes of the th Klein-Gordon equation. The Wightman function is given adopting the mode sum as shown in \eqref{wight}. Because the presence of a cosmic string does not introduce additional curvature in the spacetime for points outside the string's core, the Wightman function is expressed as the sum of two contributions, the first one  due to pure AdS spacetime, and the second is due to the presence of cosmic string in this background as expressed in \eqref{wightman-function-expanded}. Due to this fact the VEV of field squared developed in \ref{Field_squared}, is also expressed as the sum a pure AdS contribution plus a correction associated with the cosmic string. The VEV of the field squared in AdS space has been extensively analyzed in literature, our main objective here is to study the cosmic string's  contribution, $\langle|\varphi|^2\rangle_{cs}$, whose result is presented in  \eqref{phi-squared}. By using this expression we provide some asymptotic behavior considering  points near and far the string. Also we provide in Fig.\ref{fig1_0} the behavior of $\langle|\varphi|^2\rangle_{cs}$  as a function of $r/w$, that is the proper distance from the string in units of AdS curvature radius.  In subsection \ref{Energy-momentum} we develop the calculation of the VEV of the energy-momentum tensor  associated with the cosmic string, $\left\langle T_{\mu }^{\mu }\right\rangle _{\mathrm{cs}}^{\mathrm{AdS}}$. Its final expression is given by \eqref{E-M-cosmic-string}. We have checked that this tensor satisfies the trace identity, Eq. \eqref{EM-trace}, and conservation condition expressed by Eq.s \eqref{c-eq-1} and \eqref{c-eq-2}. Moreover, we present in Fig. \ref{fig4} the behavior of the energy density induced by the presence of a  cosmic string geometry, $\langle T_{0}^{0}\rangle_{\rm{cs}}$, as function of $r/w$, considering $D=4$ and different values of the parameters associated with the system.

The section \ref{sec3} is devoted to the analysis of fermionic vacuum. In subsection \ref{DiracEq} we obtain the complete set of normalized positive/negative solution of the Dirac equation in $(1+4)-$dimensional AdS space in the presence of cosmic string. In order to do that we used a specific representation for the Dirac matrices. By developing the summation over the complete set of quantum numbers associated with the fermionic modes,  we present the formal expression for the condensate fermionic, Eq. \eqref{FC2}. By a straightforward procedure we have shown that this expression can be decomposed in a sum of two contributions, the first due to the AdS space, and the other induced by the string (see Eq. \eqref{FC_CS_Ren}). The expression obtained for cosmic string induced contribution, $\langle \bar{\psi}\psi \rangle _{\mathrm{cs}}$, is presented in \eqref{FC-cs}. Its behavior as function of $r/w$ is presented in figure \ref{fig1a}. In subsection \ref{Fermionic_energy-momentum} we calculate the VEV of the energy-momentum tensor. We also have shown that this observable can be decomposed in the form presented in \eqref{decomposition}. The contribution induced by the cosmic string, $\left\langle T_{\mu }^{\mu }\right\rangle _{\mathrm{cs}}$ is given in \eqref{Tllcs}. For this fermionic case we have also proved that this tensor obeys the conservation condition and its trace satisfies relation $\left\langle T_{\mu }^{\mu }\right\rangle _{\mathrm{cs}}=sm\left\langle \bar{\psi}\psi \right\rangle _{\mathrm{cs}}$. The figure \ref{fig1_a} exhibits the behavior of the energy density, $\left\langle T_{0}^{0}\right\rangle _{\mathrm{cs}}$, as function of $r/w$ for different values of $q$.

We would like to finish this section by saying that all the plots that present the behavior of any observable as function of $r/w$ diverges for points near the string. This fact is consequence of the idealized model adopted  to the cosmic string. A more realistic model, presenting a non-vanishing radius for the string could prevent this problem; moreover, also all plots shown that the intensity of any observable increases when we increase the parameter $q$ associated with the planar angle deficit for the hyper-surface orthogonal to the string.

\section*{Acknowledgments}
I would like to thank Aram. A. Sharian, Herondy F. Santana Mota, Stefano Bellucci and Wagner Oliveira dos Santos. This work is partially supported by Brazilian Agency Council for Scientific and Technology Development (CNPq) under Grant n$\textsuperscript{\underline{o}}$ 304332/2024-0.

\end{document}